\documentclass[aps,pre,amsmath,amssym,amsthm,superscriptaddress,reprint, nofootinbib]{revtex4-2}
\usepackage[colorlinks=true,urlcolor=blue,citecolor=blue,linkcolor=blue]{hyperref}

\usepackage{graphicx}
\usepackage{dcolumn}
\usepackage{kotex}
\usepackage{natbib}
\usepackage{color}
\usepackage{amsfonts}
\usepackage{hyperref}
\usepackage{algorithm}
\usepackage{algcompatible}
\usepackage{newfloat}
\usepackage{makecell}
\usepackage{bm}

\usepackage{setspace}
\usepackage{multirow}
\usepackage{sidecap}
\usepackage{enumerate}

\definecolor{mygray}{gray}{0.5}

\begin{document}

\title{Anomaly, class division, and decoupling in income dynamics}

\author{Jaeseok Hur}
\affiliation{Department of Physics, Korea Advanced Institute of Science and Technology, Daejeon 34141, Korea}

\author{Meesoon Ha}
\email[Contact author:~]{msha@chosun.ac.kr}
\affiliation{Department of Physics Education, Chosun University, Gwangju, 61452, Korea}

\author{Hawoong \surname{Jeong}}
\email[Contact author:~]{hjeong@kaist.edu}
\affiliation{Department of Physics, Korea Advanced Institute of Science and Technology, Daejeon 34141, Korea}
\affiliation{Center of Complex Systems, KAIST, Daejeon 34141, Korea}

\date{\today}

\begin{abstract}
Economic inequality emerges from the interplay between regional growth-rate differences and the interaction network that couples regions. We propose a minimal income-dynamics model, where heterogeneity is governed by growth-rate assortativity $\mathcal{A}$ and regional concentration $\mathcal{R}$, allowing us to quantify the spatiotemporal patterns of empirically observed log-income distributions. To systematically analyze these patterns, we derive closed-form approximations for the Hellinger distance and the Gini index in limiting configurations. Our findings highlight the spatial segregation of growth rates as a key driver of economic class division and demonstrate how small-world shortcuts in the underlying network can disrupt this segregation. Finally, our framework provides a robust explanation for the bimodality and strong regional correlations found in global income distributions.

\end{abstract}

\maketitle

\section{Introduction}  

Since the publication of Piketty's {\it Capital in the Twenty-first Century}~\cite{piketty2014capital} ignited a debate on the mechanism of economic inequality, both wealth and income distributions have been widely addressed in both theoretical~\cite{gibrat1931inegalits,pestieau1979model,pestieau1982model,chotikapanich2008modeling,chakrabarti2013econophysics,nirei2016pareto,gabaix2016dynamics} and empirical studies~\cite{piketty2003income,bourguignon2002inequality, milanovic2015global,milanovic2024three,liberati2015world,milanovic2013global, sala2006world,pinkovskiy2009parametric,anand2008we,van2014changing,lakner2016global,owid-the-history-of-global-economic-inequality}. In {\it World Inequality Report 2022}~\cite{chancel2022world}, global income inequality for the period 1950--2010 was attributed to regional location~\cite{bourguignon2002inequality,milanovic2015global}. Such inequality, as represented by the Gini index, peaked in the late twentieth century, and income distributions were bimodal.
Strong regional correlations were observed in the global income levels of countries~\cite{roser2023income}, and the corresponding regional convergence was studied in~\cite{rey1999us,yamamoto2008scales}.

The most recent study by Milanovic~\cite{milanovic2024three} divided the history of global-income inequality into three eras: In the first era, both within- and between-inequality increases, characterized by persistently increasing segregation of income levels. In the second era, high global inequality and regional segregation become chronic, characterized by the bimodality of income distributions. In the third era, corresponding to the contemporary era, it no longer exhibits any bimodality in income distributions, caused by the acceleration of growth rates in developing countries, such as China and India~\cite{milanovic2013global,liberati2015world,milanovic2024three}. 

In this paper we provide a comprehensive picture of inequality in the first two eras 
of~\cite{milanovic2024three}, where spatiotemporal patterns of income dynamics exhibit bimodality (class division between the rich and the poor) and regional segregation (location dependence) of income levels~\cite{milanovic2015global}. We address fundamental questions in economic inequality: how can such a mechanism be formulated, in the context of a prototype pedagogical model, and what are the key ingredients of such a model? 
Modeling income dynamics as a binary mixture of growth rates, which is analogous to quenched disorder in physical systems, we implement both class division and decoupling of two growth rate groups. 
Our hypothesis is that bimodality and regional segregation of income levels are driven by heterogeneity of growth rates and low connectivity between regions with different growth rates. Hence, the model is tested in a one-dimensional (1D) ring analytically and in a sparse regular and small-world (SW) network numerically.

\begin{figure}[]
\includegraphics[width=\columnwidth]{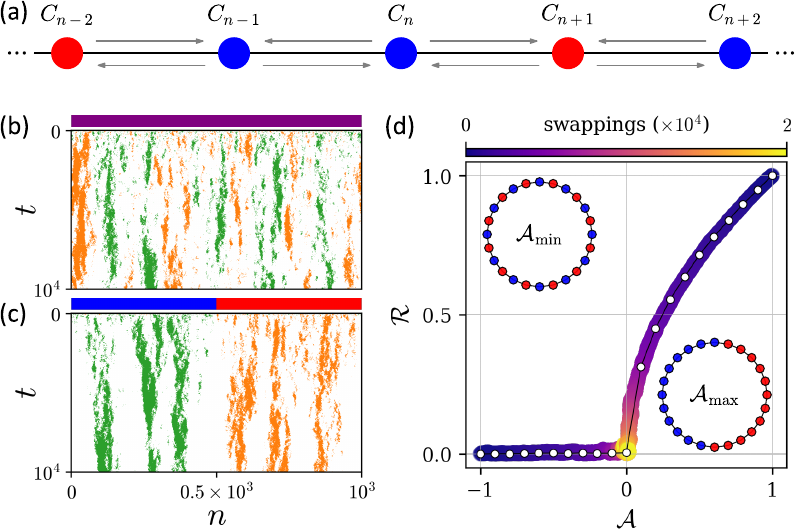}
    \caption{(a) Schematic illustration of income dynamics with a binary mixture of regional growth rate in the HBM model: ${\color{red}\bullet}~(\alpha_{+},
    \text{red})$ and ${\color{blue}\bullet}~(\alpha_{-}, \text{blue})$ for $\alpha_\pm=\alpha\pm \Delta \alpha$ and income ($C$) transfer (either $\rightarrow$ or $\leftarrow$) between two nearest-neighboring sites. 
    (b, c) Snapshots of spatiotemporal patterns for a top-rich/bottom-poor 10\% (orange/green) class are taken from a single run for two extreme configurations: ($\mathcal{A}_{\min},\mathcal{A}_{\max}$): The position index $n$ is shown in horizontal from left to right, and time $t$ is in vertical from top to bottom. Here $N=10^3,\alpha=10^{-2},\Delta\alpha=10^{-3},\beta^2=10^{-3},~\mbox{and}~ J=10^{-1}$ in Eq.~\eqref{eq-HBM}. (d) Random pair swapping trajectories of $(\mathcal{A}, \mathcal{R})$. A path starts from two extreme configurations with $(\mathcal{A}_{\rm min/max},\mathcal{R}_{\rm min/max})$ (see insets). Each random pair-swapping trial is represented by color gradation. The interval of simulation samples ($\circ$) is 0.1 in [$\mathcal{A}_{\rm min},\mathcal{A}_{\rm max}$].}
    \label{fig1-model}
\end{figure}

Our model can be considered as the heterogeneous case of the wealth-dynamics model by Bouchaud and M{\'e}zard (BM)~\cite{bouchaud2000wealth}, in which the configuration of binary growth rates is tuned by regional growth rate assortativity $\mathcal{A}$ and concentration $\mathcal{R}$ (see Fig.~\ref{fig1-model}). For the homogeneous case, we provide rigorous analytical derivations and more detailed results in \textit{Appendix A} and Sec.~I in the Supplemental Material (SM)~\cite{SM}.

Since the allocation of binary growth rates influences the spatiotemporal patterns of income dynamics, we unveil the conditions under which bimodal income distributions can be observed. 
In a 1D ring, a slowly decaying field exponent $\eta$ produces sub-diffusive broadening of log-income within groups, while configuration governs structure: $\mathcal{R}$ drives decoupling and bimodality, and $\mathcal{A}$ sets the pace of inequality growth. 
Closed-form approximations for the Hellinger distance $h$ and the Gini index $g$ quantify these effects. Adding SW shortcuts weakens long-range correlations, disrupts segregation, and collapses bimodality, identifying spatial segregation, not growth rate differences alone, as a key mechanism of class division.
Finally, we conclude with a comprehensive picture of global income inequality and a possible remedy to alleviate it.

\section{Model}

We introduce a binary mixture of growth rates into the BM model on a network topology. The following stochastic differential equation (SDE) is for the heterogeneous BM (HBM) model [see Fig.~\ref{fig1-model}(a)]: 
\begin{align}
    dC_n=\alpha_n C_ndt+\beta C_ndW_{t,n}+J(\bar{C}_n^{(k)}-C_n)dt,
    \label{eq-HBM}
\end{align}
where $C_n$ is the income at a node $n$, $dt$ is a time interval, $W_{t,n}$ is the Wiener process at $n$ at time $t$, $\alpha_n=\alpha\pm\Delta\alpha$ is the growth rate at $n$, $\beta$ is volatility, $J$ is the strength of interaction between two coupled neighbors, and $\bar{C}_n^{(k)}$ is the average income for $k$ neighbors at $n$, respectively. 
Initially, all nodes have the same income. This SDE follows the It{\^o} interpretation. We note that both $\alpha$ and $\Delta\alpha$ are positive constants, preserving $\alpha>\Delta\alpha$.

Since the SDE [Eq.~\eqref{eq-HBM}] incorporates quenched disorder, the growth-rate configuration remains static. The disorder of the growth rate configuration disrupts the translational invariance of the system. The transformation analysis of the ordinary BM model cannot hold.

Without loss of generality, we consider the case that the number of nodes for the group of $\alpha_+$ is the same as that of $\alpha_-$: $N_+=N_-$ and $\alpha_{\pm}=\alpha \pm \Delta \alpha$ [see Fig.~\ref{fig1-model}(b) and (c) for two extreme configurations]. In Fig.~\ref{fig1-model}(b) the alternative allocation generates the spatiotemporal patterns of top-rich (orange)/bottom-poor (green) 10
while in Fig.~\ref{fig1-model}(c), the fully separated allocation displays regional segregation (class division) and long-range spatial correlations (clustering). 
\subsection*{Statistical properties and control parameters}

We quantify the statistical properties of the allocation of $\alpha_\pm$ and generate all possible configurations in terms of two relevant control parameters.

The first parameter of Fig.~\ref{fig1-model}(d), growth-rate assortativity $\mathcal{A}$, quantifies the connectivity between regions with similar $\alpha$ values connected in a given network:
\begin{align}
    \mathcal{A}\equiv\frac{{\rm Cov}(\alpha,\alpha')}{\sqrt{{\rm Var}(\alpha){\rm Var}(\alpha')}},
    \label{eq-assortativity}
\end{align}
where $\alpha$ and $\alpha’$ are the growth rates in two connected regions for a given network, respectively. 

For the binary mixture in the 1D ring with $N_+=N_-$, $\mathcal{A}=\rho^{(1)}-\rho^{(2)}$, where $\rho^{(1)}$ and $\rho^{(2)}$ are homogeneous and heterogeneous link densities, respectively (see Sec.~II~A~in SM~\cite{SM} for the definition and detailed derivations). The actual bounds~\cite{Pearson} of $\mathcal{A}$ depend on the network topology~\cite{cinelli2020network}. 
In the 1D ring, the alternative allocation of growth rates becomes the lower bound of $\mathcal{A}$, and the fully separated allocation becomes the upper bound of $\mathcal{A}$: $\mathcal{A}_{\rm min} = -1$ for Fig.~\ref{fig1-model}(b) and $\mathcal{A}_{\rm max}=+1-4/N$ for Fig.~\ref{fig1-model}(c), illustrated as the insets in Fig.~\ref{fig1-model}(d).

The second parameter in Fig.~\ref{fig1-model}(d), $\mathcal{R}$, is derived from the phase order parameter in the Kuramoto model~\cite{kuramoto1975international}, which represents the polar concentration of growth rates in the 1D ring. Intuitively, $\mathcal{R}$ serves as a topological metric that quantifies the spatial clustering of growth rates on the ring as well. Thus, $(r_\pm,\psi_\pm)$ is $r_\pm e^{i\psi_{\pm}}=\frac{1}{N_\pm}\sum_{j}e^{i\phi_j^{(\pm)}}$, 
where $\phi_j^{(\pm)}$ is the angular argument for binary growth rate groups, respectively. 

In the 1D HBM model, all regions are assigned in a 1D ring with the same intervals. Thus, $\phi_j=2\pi m/N$, where $m\in\{0,\dots,N-1\}$. For the case of $N_+=N_-$ and $r_\pm \ne 0$, two important quantities become $r_+=r_-=r$ and $\Delta\psi=|\psi_+ -\psi_-|=\pi$.
For any configuration, $\Delta\psi$ is either $\pi$ or ``not defined" ($r=0$), so we only consider $r$. For the perfectly disassortative case with $\mathcal{A}_{\rm min}$ and the perfectly assortative case with $\mathcal{A}_{\rm max}$, $r\to0$ and $r\to 2/\pi$ as $N\to\infty$, which are the minimum and the maximum, respectively (see Sec.~II~B and Fig.~S5 in SM~\cite{SM} for detailed definitions and discussions). Hence, $\mathcal{R}$ is defined as a statistical control parameter: 
\begin{align}
\mathcal{R}\equiv r/(2/\pi).
 \label{eq-concentration}
\end{align}

\section{Results}

To characterize spatiotemporal patterns in income distributions, we employ two standard statistical measures: Hellinger distance $h$~\cite{hellinger1909neue} and Gini index $g$~\cite{gini1912variabilita}.

\begin{figure}[]
\includegraphics[width=\columnwidth]{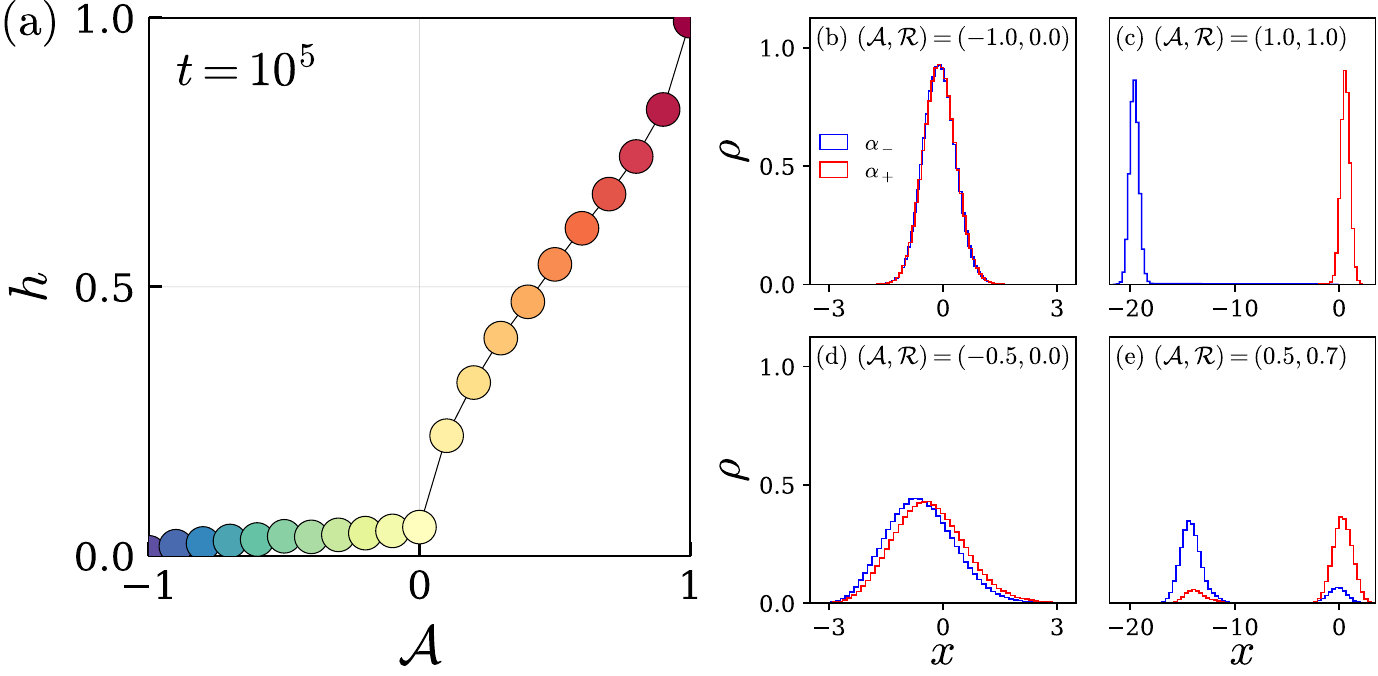}
    \caption{Configuration effect on decoupling by Hellinger distance $h$: (a) $h$ in Eq.~\eqref{eq-h} plotted at $t=10^5$ against $\mathcal{A}$ with $\mathcal{R}(\mathcal{A}$); see Fig.~\ref{fig1-model}(d). Selected snapshots for unimodal and bimodal distributions: (b), (c), (d), and (e) are the cases of $(\mathcal{A}, \mathcal{R})=(-1.0, 0.0),~(-0.5,0.0),~(0.5,0.7),~\mbox{and}~(1.0,1.0)$ for $t=10^4$, respectively. Here, $N=10^4$, results are obtained by 128 runs, and the other parameters are the same as Fig.~\ref{fig1-model}.}
    \label{fig2-h}
\end{figure}

First, $h$ is defined as
\begin{align}
h(\rho_1,\rho_2)\equiv\sqrt{\frac{1}{2}\int\left(\sqrt{\rho_1(x)}-\sqrt{\rho_2(x)}\right)^2dx},
    \label{eq-h}
\end{align}
where $0\leq h\leq 1$ and $\rho_{i}(x)$ is a probability distribution of log-normalized income $x$~\cite{log-income}. This follows the Lebesgue measure. The entire $\rho$ can be decomposed by $\rho_{\alpha_{\pm}}$ (see Fig.~\ref{fig2-h}), $\rho=(\rho_{\alpha_-}+\rho_{\rm \alpha_+})/2$. 
Hence $h(\rho_{\alpha_-},\rho_{\alpha_+})=0~(1)$ is a perfectly coupled (decoupled) state (see Figs.~\ref{fig1-model} and~\ref{fig2-h}).
%

Second, $g$ is defined as
\begin{align}
    g\equiv\frac{1}{2\langle{c}\rangle}\int_{0}^{\infty}\int_{0}^{\infty}\rho(c)\rho(c')|c-c'|dcdc',
    \label{eq-g}
\end{align}
where $0\leq g\leq 1$ and $\rho(c)$ is a probability distribution of 
normalized income $c\equiv C/\bar{C}$. Hence, $g=0~(1)$ is the perfect equality (extreme inequality as a condensation state with the whole income monopolized by a few regions).

For the sampling of growth rate configurations, we perform the random pair swapping algorithm: (1) Start from two extreme configurations, $(\mathcal{A}_{\rm min},\mathcal{R}_{\rm min})$ and $(\mathcal{A}_{\rm max},\mathcal{R}_{\rm max})$, respectively. (2) Select a random pair of regions and switch positions. If this process is repeated for large iterations, $(\mathcal{A},\mathcal{R})$ almost converges to near $(0,0)$ [see Fig.~\ref{fig1-model}(d)].
To elucidate the impact of growth rate configurations of regional growth rates on income dynamics at a glance, we compare two extreme cases of the configurations, $\mathcal{A}_{\rm min}$ and $\mathcal{A}_{\rm max}$: 
For the perfectly disassortative case, $\rho(x,t;\mathcal{A}_{\min})$ is
\begin{align}
\rho(x,t;\mathcal{A}_{\min})\approx\rho_{\alpha_\pm}(x,t)\sim\mathcal{N}\left(\mu_t,\sigma_t^2\right),
    \label{dist-AA-1}
\end{align}
where $\sigma_t^2=\beta^2t^{\lambda}/(2Ja_0)$, $\mu_t=-\sigma_t^2/2$ for large $t$, and $0.5\leq \lambda\leq 1$. These results are consistent with the homogeneous BM model in the 1D ring topology (see Sec.~I in SM~\cite{SM} for detailed derivations). For the case of $h\approx 0$, $\rho_{\alpha_\pm}$ almost perfectly overlap each other [see Fig.~\ref{fig2-h}(b)]. The perfectly disassortative configuration neutralizes the impact of heterogeneous growth rates on income distributions. 
For the perfectly assortative case, $\rho(x,t;\mathcal{A}_{\max})$ for large $t$ is divided into three parts [see Fig.~\ref{fig2-h}(c)] with two well-separated Gaussian peaks: (1) Gaussian peak around the first mode (head), $\rho^{\rm (h)}$, (2) flat region between the two peaks (body), $\rho^{\rm (b)}$, and (3) Gaussian peak around the second mode (tail), $\rho^{\rm (t)}$:
\begin{align}
    \rho(x,t;\mathcal{A}_{\max})\approx
    \begin{cases}
    \rho^{(\rm h)}(x,t) &\sim \mathcal{N}(\mu_t^{-},\sigma_t^2), \\
    \rho^{(\rm b)}(x,t) &\sim {\rm const}, \\
    \rho^{(\rm t)}(x,t) &\sim \mathcal{N}(\mu_t^{+},\sigma_t^2),
    \label{dist-AA1}
    \end{cases}
\end{align}
where $\mu_t^{\pm}\approx\mu_t+\ln\left[2/(1+e^{\mp2\Delta\alpha t})\right]$. $\sigma_t^2$ and $\mu_t$ are the same as before. $\rho^{\rm (h)}$ and $\rho^{\rm (t)}$ are formed by lower and higher growth rate nodes, respectively (see Sec.~III and Figs.~S6 and S7 in SM~\cite{SM} for details). 

While the motions in the same growth rate group are subdiffusive, represented by $\sigma_t^2$, the relative motion between the different groups of $\alpha_{\pm}$ is ballistic, represented by income level segregation that increases linearly in $t$, $\Delta\mu=\langle{x_{\alpha+\Delta\alpha}}\rangle-\langle{x_{\alpha-\Delta\alpha}}\rangle\approx2\Delta\alpha t$. Since $\Delta\mu$ increases faster than $\sigma_t$, there is almost no overlap between $\rho_{\pm}$ for large $t$ and it demonstrates $h\approx1$. In other words, the perfectly assortative configuration maximizes decoupling. Surprisingly, this linearity is also consistent for {\it Path 2} configurations such that $\Delta\mu\approx\mathcal{A}\times2\Delta\alpha t$. It exhibits that configurational property $\mathcal{A}$ also controls income level segregation $\Delta\mu$ (see Sec.~III and Figs.~S8 and S9 in SM~\cite{SM} for detailed derivations and numerical confirmations).
%
\begin{figure}[b]
\includegraphics[width=\columnwidth]{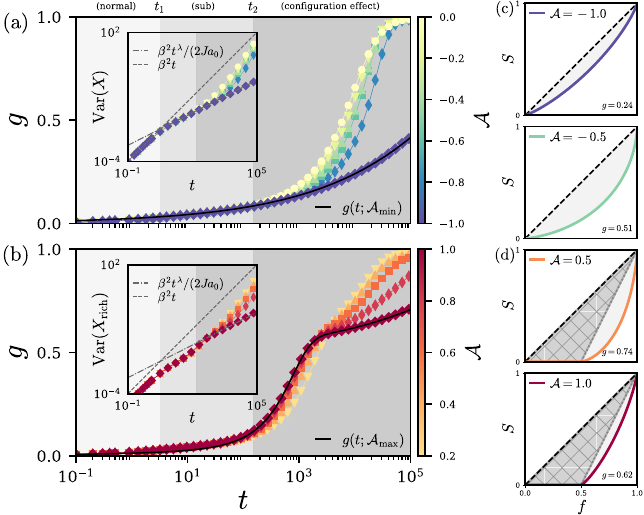}
    \caption{Configuration effect on inequality by Gini index $g$ against $t$: Each color reflects the selected samples of $\mathcal{A}$ with $\mathcal{R}$ in Fig.~\ref{fig1-model}~(d). {\it Path 1} (a) of $(\mathcal{A}_{\min}, \mathcal{R}_{\min})\to (0,0)$ and {\it Path 2} (b) of $(\mathcal{A}_{\max}, \mathcal{R}_{\max})\to (0,0)$, respectively. As $t$ elapses, the system exhibits normal diffusion (below $t_1$), subdiffusion ($t_1\le t\le t_2$) and finally reaches the configuration-effect dominant diffusion (above $t_2$). Insets in (a) and (b) are ${\rm Var}(X)$ and ${\rm Var}(X_{\rm rich})$, respectively, where the guided lines are drawn for normal diffusion and subdiffusion as described, respectively. Lorenz curves (c) and (d) of $S$ against $f$ at $t=10^4$ correspond to Figs.~\ref{fig2-h}(b)-\ref{fig2-h}(d) and Figs.~\ref{fig2-h}(c)-\ref{fig2-h}(e), where plain and hatched shadow regions illustrate the contributions of class division and diffusion to $g$, respectively.}
    \label{fig3-g}
\end{figure}

For the perfectly disassortative case [see Fig.~\ref{fig3-g}(a)], 
$g(t;\mathcal{A}_{\min})$ is
\begin{align}
    g(t;\mathcal{A}_{\min})\approx{\rm erf}(\sigma_t/2).
    \label{gini-Amin}
\end{align}
Increasing $g(t)$ is only driven by sub-diffusion because there is almost no decoupling between $\rho_{\alpha_\pm}(x,t)$.

For the perfectly assortative case [see Fig.~\ref{fig3-g}(b)], $g(t;\mathcal{A}_{\max})$ is
\begin{align}
    g(t;\mathcal{A}_{\max})\approx\frac{1}{2}\left(1-\frac{2}{1+e^{2\Delta\alpha t}}\right)+\frac{1}{2}{\rm erf}\left( \sigma_t/2 \right).
    \label{gini-Amax}
\end{align}
The first term of Eq.~\eqref{gini-Amax} captures the inequality between the $\alpha_\pm$ groups. It converges rapidly to $1/2$ because the regional segregation of income levels leaves the $\alpha_{-}$ group (comprising 50\% of the population) with a negligible share of the total income. The second term accounts for the inequality within the $\alpha_{+}$ group, converging gradually to $1/2$ due to the subdiffusive broadening of the Gaussian peaks shown in Fig.~\ref{fig2-h}(c). Detailed derivations are provided in \textit{Appendix B} and Sec.~III~B in SM~\cite{SM}. This result serves as a compelling illustration of the dominant influence of location (region or country)~\cite{milanovic2015global} on income levels, highlighting the distinct contributions of between- and within-group inequalities~\cite{milanovic2013global} to the global system.

In Fig.~\ref{fig3-g} we present the time evolution of $g$ across various $\alpha$ configurations. For all cases, the dynamics exhibit three distinct regimes: (1) Normal diffusion, (2) Gaussian sub-diffusion, and (3) non-Gaussian diffusion [see insets of Figs.~\ref{fig3-g}(a) and \ref{fig3-g}(b)]. The transitions between these regimes occur at timescales $t_1=\left[2Ja_0\right]^{1/(\lambda-1)}$ and $t_2=\left[2Ja_0\Delta\alpha^2/\beta^2\right]^{1/(\lambda-2)}$, marking the crossovers from sub-diffusion to normal diffusion and subsequently to ballistic or superdiffusive motion. These crossovers are highlighted by guidelines and shaded regions (see Sec.~III~D and Fig.~S10 in SM~\cite{SM} also for detailed derivations and numerical confirmations). Unlike the first two regimes, which are characteristic of the homogeneous BM model, the third regime is unique to the HBM model, arising specifically from configuration effects.

Along \textit{Path 1}, $\rho_{+}$ and $\rho_{-}$ overlap significantly, resulting in a single peak for the total distribution $\rho(x)$ [see Fig.~\ref{fig2-h}(b) and \ref{fig2-h}(d)]. Consequently, $g$ evolves on a single time scale, as described by Eq.~\eqref{gini-Amin}, with inequality driven solely by diffusion. In this regime, $g$ increases monotonically with $\mathcal{A}$ [see Fig.~\ref{fig3-g}(a)].
Along \textit{Path 2}, $\rho_{+}$ and $\rho_{-}$ separate, leading to a bimodal $\rho(x)$ [see Figs.~\ref{fig2-h}(c) and \ref{fig2-h}(e)]. Consequently, inequality is governed by both class division and diffusion. As indicated by Eq.~\eqref{gini-Amax}, the substantial income gap between the two peaks contributes a baseline value of $0.5$ to $g$. This manifests as a horizontal segment in the Lorenz curve, implying a kind of economic extinction of the poor class [see Figs.~\ref{fig3-g}(c) and \ref{fig3-g}(d)]. 
The remaining contribution to $g$ arises from diffusion within the rich class. We therefore analyze ${\rm Var}(X_{\rm rich})$, defined for incomes $X_{\rm rich}$ exceeding the median. In this case, decreasing $\mathcal{A}$ enhances ${\rm Var}(X_{\rm rich})$, driving the system toward a super-diffusive regime that accelerates condensation ($g\to 1$) [see Fig.~\ref{fig3-g}(b) and Sec.~III~D in SM~\cite{SM} for details]. Remarkably, introducing even modest heterogeneity and mixing within the 1D ring topology induces super-diffusive behavior, a phenomenon absent in the homogeneous BM model.

\begin{figure}[]    
\includegraphics[width=\columnwidth]{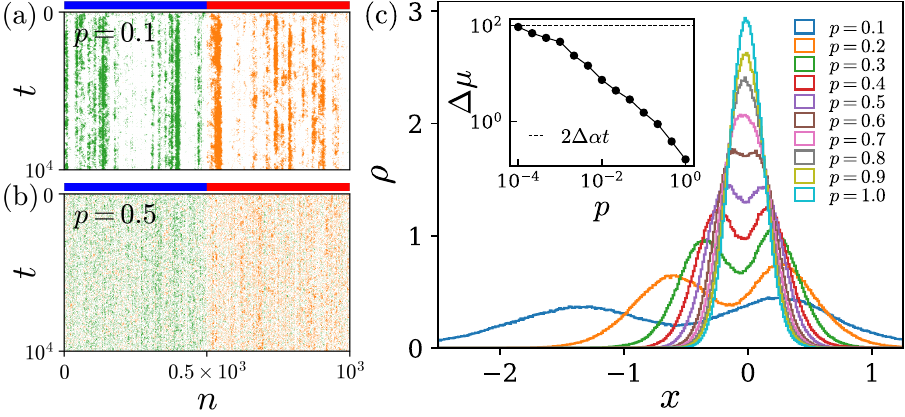}
    \caption{SW effect on the HBM model: Spatiotemporal patterns of top-rich (orange)/bottom-poor (green) 10\% classes over positional index $n$ for a fully separated configuration with rewiring probability (a) $p= 0.1$ and (b) $p=0.5$ for adding shortcuts. (c) Snapshots of log-income distributions for various $p$ values at $t=10^4$. The inset displays income-level segregation $\Delta\mu$ against $p$. Here we employ the WS network with $k=4$, and $N=10^3,\Delta\alpha=10^{-3}$ in (a) and (b); $N=10^4,\Delta\alpha=5\times10^{-3}$ in (c). The other parameters are the same as Fig.~\ref{fig1-model}~(b) and~(c).}
    \label{fig4-sw}
\end{figure}

Finally, we investigate the effect of small-worldness~\cite{BM-WS} on income distributions in the HBM model (see Fig.~\ref{fig4-sw} and Sec.~IV, and Figs.~S11-S13 in SM~\cite{SM} for detailed numerical confirmations). We employ the Watts-Strogatz (WS) network, initializing the growth-rate configuration for the case of $\mathcal{A}_{\max}$ prior to rewiring with probability $p$. This rewiring process introduces heterogeneous links with probability $p/2$, thereby reducing assortativity such that $\mathcal{A}\simeq1-p$. Increasing $p$ establishes shortcuts between spatially segregated regions with high and low growth rates. In contrast to the homogeneous BM model, small-world (SW) shortcuts in the HBM model disrupt not only long-range correlations but also regional income segregation, effectively dismantling log-income bimodality. As $p$ increases, rich and poor clusters shrink, and long-range correlations decline [see Fig.~\ref{fig4-sw}(a) and~(b)]. Moreover, regional income segregation weakens as poor (rich) clusters emerge within high (low) growth-rate regions. Consequently, the segregation of income levels $\Delta\mu$ decreases, and bimodality collapses as $p$ rises [see Fig.~\ref{fig4-sw}(c) and Sec.~IV in SM~\cite{SM} for detailed numerical verification].

\section{Summary and Remarks} 

To sum up, we investigated how regional growth-rate configurations affect income distributions. The emergence of bimodality, spatial correlations, and regional segregation—features consistent with global income patterns over the past half-century—can be linked to the absence of small-world connectivity in sparse regular networks, corresponding to small $\eta$ (see \textit{Appendix A}). We found that regional concentration of growth rates ($\mathcal{R}$) is associated with the development of bimodality and income-level segregation, whereas growth-rate assortativity ($\mathcal{A}$) influences the diffusive behavior and long-term inequality.

For Watts–Strogatz (WS) topologies, the introduction of small-world shortcuts connects high- and low-growth regions, which tends to reduce long-range correlations and weaken the bimodal structure of the income distribution. The reduced diffusion across growth-rate interfaces bears some similarity to localization effects induced by quenched disorder~\cite{abanin2019colloquium,nandkishore2015many}, while the resulting segregation of income levels is reminiscent of phase-separation-like behavior observed in nonequilibrium active matter systems~\cite{cates2015motility,fily2012athermal}.

Our model can be taken as a possible framework for interpreting aspects of the historical evolution of global inequality~\cite{chancel2022world,owid-the-history-of-global-economic-inequality,iref-reducing-global-inequality}, particularly the first two of the three eras identified by Milanovic~\cite{milanovic2024three}. Sparse regular networks with large $\mathcal{R}$ are consistent with the first era, in which both within- and between-region inequalities increase. WS networks with small $p$, which exhibit stationary bimodal distributions, may be related to the second era, where global inequality and regional segregation become persistent. The third (contemporary) era, which does not show clear bimodality, has been associated with accelerated growth in developing economies such as China and India~\cite{milanovic2013global,liberati2015world,milanovic2024three}. Incorporating temporal growth rates, $\alpha(t)$, would be a natural extension, it is beyond the scope of the present work. In addition, recent studies suggesting that trade and migration contribute to reducing global inequality~\cite{hammar2020global,iref-reducing-global-inequality} are qualitatively consistent with our results for WS networks with sufficiently large $p$.

A natural direction for future work is to consider continuous spectra of growth rates. While normally distributed growth rates have been studied in noninteracting~\cite{pluchino2018talent} and mean-field limits~\cite{bernard2025mean}, their behavior on complex networks remains to be explored in more detail~\cite{hur2024interplay}. The role of heterogeneous volatility also remains an open question. Long-range correlations are not limited to income, but have been reported in other social indicators, such as housing prices~\cite{becharat2025diffusive} and voting patterns~\cite{fernandez2014voter}. In systems with spatial heterogeneity in local conditions, including growth rates or infrastructure, the metrics ($\mathcal{A}, \mathcal{R}$) could serve as one possible measure to characterize configuration effects. Finally, capturing the full history of global income inequality may require extensions to temporal networks; in this context, the evolving structure of the world trade network~\cite{serrano2007patterns,cha2010patterns} could offer additional insight.
\begin{acknowledgments} 
This research was supported by Basic Science Research Program through the National Research Foundation of Korea (NRF) (KR) [NRF-RS-2025-00514776~(J.H. and H.J.) and NRF-RS-2026-25489888(J.H. and M.H.)].
\end{acknowledgments} 

\section*{DATA AVAILABILITY}
The data for the networks used in this paper are publicly available in the cited references. The data supporting the figures are not publicly available but can be provided by the authors upon reasonable request.

\begin{appendix}

\setcounter{equation}{0}
\setcounter{table}{0}

\renewcommand{\theequation}{A\arabic{equation}}
\renewcommand{\thetable}{A\arabic{table}}

\section*{Appendix A: Homogeneous BM model on sparse regular networks} 

In general, the original Bouchaud and M{\'e}zard (BM) model~\cite{bouchaud2000wealth} is represented by a stochastic differential equation (SDE):
\begin{align}
    dC_n=\alpha C_ndt+\beta C_ndW_{t,n}
    +\sum_{m(\neq n)}\!(J_{nm}C_m-J_{mn}C_n)dt,
    \label{eq-BM}
\end{align}
where $n$ is an agent index, $C_n$ is the income of node $n$, $dt$ is a time interval, $W_{t,n}$ is the Wiener process of $n$ at time $t$, $\alpha$ is the growth rate, $\beta$ is the volatility and $J_{nm}$ is the element of an interaction matrix $\bold J$, respectively. This follows the It{\^o} interpretation.

To focus on the relative income of the nodes, we denote 
a normalized income $c\equiv C/\bar{C}$,
where $\bar{C}$ is an average income. Then, we can rewrite Eq.~\eqref{eq-BM} with the proper notation~\cite{SDE-normalizedwealth} as
\begin{align}
    dc_n=J\sum_{\{m|a_{nm}=1\}}\left(\frac{c_m}{k_m}-\frac{c_n}{k_n}\right)dt+\beta c_ndW_{t,n},
    \label{eq-BM-c}
\end{align}
where $a_{mn}$ is the element of the adjacency matrix $\bold a$ for an interaction network, and the strength of the interaction between node $n$ and node $m$ is $J_{nm}=J/k_m$ with $J>0$, and the degree of node $m$ is $k_m$. 
For the case of random networks, $J_{mn}=a_{mn}J/\langle{k}\rangle$ with the average degree $\langle k\rangle$. This rescaling of $C$ and the normalization by division $\bar{C}$ do not alter Eq.~\eqref{eq-BM} nor the Gini index $g$ and the Hellinger distance $h$. 
In our numerical simulation based on Eq.~\eqref{eq-BM},
this process is identical to Eq.~\eqref{eq-BM-c}, which also guaranties $\langle{c}\rangle=1$.
For a sufficiently dense (or small-world) network, $\rho(c,t)$ converges to the stationary distribution for large $t$, and most previous studies focus on it~\cite{souma2001small,souma2003wealth,garlaschelli2004wealth,garlaschelli2008effects,medo2009breakdown,ichinomiya2013power,ma2013distribution,ichinomiya2012bouchaud,ichinomiya2013power}. 

For a regular network, Eq.~\eqref{eq-BM-c} is reduced by:  
\begin{align}
    dc_n=J(\bar{c}_n^{(k)}-c_n)dt+\beta c_ndW_{t,n},
    \label{eq-c}
\end{align}
where $\bar{c}_n^{(k)}$ is the average normalized income over $k$ neighbors of a node $n$. Equation~\eqref{eq-c} is solved by the effective field theory (EFT) ansatz~\cite{ma2013distribution}
\begin{align}
    \bar{c}_n^{(k)}\to\theta(\eta)c_n^{1-\eta},
    \label{eft-ansatz}
\end{align}
where a field exponent $\eta\in(0,1]$ and a normalization factor $\theta(\eta)$ from $\langle{c}\rangle=1$. The field exponent $\eta$ displays the nonlinear effect of local interactions, approximated by $c_n$ itself. Therefore,  Eq.~\eqref{eq-c} can be rewritten as
\begin{align}
    dc_n=J[\theta(\eta)c_n^{1-\eta}-c_n]dt + \beta c_ndW_{t,n}.
    \label{eq-EFT}
\end{align}
The study by Ma {\it et al.}~\cite{ma2013distribution} demonstrates that $\eta$ converges to a constant for sufficiently large $z\equiv k/N$ ($\ge 10^{-2}$) and $\rho(c)$ is the generalized inverse-gamma distribution. 

However, for small $z$, $\eta(t)$ does not converge to a stationary solution. Hence, the temporal behavior of $\eta$ should be analyzed in the sparse regular network.
Let $x\equiv\ln c$ and assume that $\eta x$ is sufficiently small, then the first-order approximation of Eq.~\eqref{eq-EFT} becomes a time-dependent {\it Ornstein-Uhlenbeck} (OU) process~\cite{gardiner1985handbook} as 
\begin{align}
    dx_n=J\eta\theta(\eta)\left[ \left(\frac{\theta(\eta)-1}{\eta\theta(\eta)}\right)-x_n \right]dt-\frac{\beta^2}{2}dt+\beta dW_{t,n},
    \label{eq-x}
\end{align}
where $\eta$ and $\theta(\eta)$ depend on $t$.

Surprisingly, for large $t$, the variance of the process of Eq.~\eqref{eq-x} is asymptotically similar to that of the ordinary OU process: $\sigma_t^2=\beta^2/[2J\eta\theta(\eta)]$~\cite{OU}, where both $\eta$ and $\theta$ are substituted as constants. This phenomenon depends on the slow decay of $\eta(t)$ as $\rho(x,t)\approx\rho_{\rm eq}^{(\rm OU)}(x,t;\eta_t,\theta_t)$.
If Eq.~\eqref{eft-ansatz} is in good approximation, the random variables, $Y=\ln{\bar{c}^{(k)}}$ and $X=\ln{c}$, are in the linear relationship as $Y=(1-\eta)X+\ln{\theta(\eta)}$. 

According to the least square linear regression: $1-\eta={\rm Cov}(X,Y)/{\rm Var}(X)$ (see Sec.~I~A and Figs.~S1 and S2 in SM~\cite{SM}), we empirically find that $\eta$ follows the time-asymptotic power-law as
\begin{align}
    \eta(t)\sim a_0t^{-\lambda}\quad\text{for large $t$},
    \label{eta-power-law}
\end{align}
where $a_0$ is constant and $0<\lambda\leq1$ for small $z$, and $\theta(\eta)\to 1$. This supports our approximation that $\rho(x,t)\approx\rho^{\rm (OU)}_{\rm eq}(x,t;\eta_t,\theta_t)$ at each point in time is approximately Gaussian. For large $t$, the variance $\sigma_t^2=\beta^2t^{\lambda}/(2Ja_0)$ and the mean $\mu_t=-\sigma_t^2/2$ of $\langle{c}\rangle=1$, such that
$c_t\sim{\rm Lognormal}(\mu_t,\sigma_t^2)$.

For the one-dimensional (1D) ring topology under the small $\beta$ condition, the SDE for $X$ corresponds to the multidimensional OU process, and the interaction matrix is marginally stable~\cite{bouchaud2024self}. Then $\eta$ is given by
\begin{align}
    \eta(t)=\frac{1-e^{-2Jt}I_0(2Jt)}{2Jte^{-2Jt}\left[I_0(2Jt)+I_1(2Jt)\right]},
\end{align}
where $I_\ell$ ($\ell=0,1,2,\dots$) is the modified Bessel function of the first kind. This expression satisfies $a_0t^{-1/2}$ for large $t$, and is also consistent with Eq.~\eqref{eta-power-law}. The power-law decay of $\eta$ corresponds to the vanishing of nonlinearity in the effective field, which results in the convergence of income of neighboring nodes. On the other hand, the variance $\sigma_t^2$ increases over time $t$, so almost all income is condensed (localized) in narrow regions for large $t$, which corresponds to the localization in the 1D stochastic heat equation or 1D Kardar-Parisi-Zhang equation as well.  

The $\ell$-ranged covariance is as
\begin{align}
    {\rm Cov}(X_n,X_{n+\ell})=\beta^2\int_{0}^{t}e^{-2Js}I_{\ell}(2Js)ds,
    \label{eq-Cov}
\end{align}
for the same time $t$, the shorter the distance $\ell$, the larger the covariance. The results in strong spatial correlations of $X$ that make the clustering of rich and poor regions, as shown in Fig.~\ref{fig1-model} (b) and (c), respectively [see Sec.~I~A in SM~\cite{SM} for details of ${\rm Var}(X)$, ${\rm Cov}(X,Y)$, $\eta(t)$, and long-range correlations]. The BM model in a sparse regular network exhibits strong spatial correlations, while that in other network cases
(the non-interactive case and the mean-field case) does not (see Fig.~S3 in SM~\cite{SM}) but log-income distributions are still unimodal.
The log-normality of $\rho(c)$ for small $z$ reported by Souma {\it et al}.~\cite{souma2001small}, is derived with a time-dependent OU process of $x\equiv\ln{c}$. We show that it is valid only for small $\beta$ and $z$.

In short, we develop a time-dependent EFT ansatz and derive temporal behaviors of statistical properties for the case of the 1D ring topology and the case of sparse regular networks by analytical and numerical.

\setcounter{equation}{0}
\setcounter{table}{0}

\renewcommand{\theequation}{B\arabic{equation}}
\renewcommand{\thetable}{B\arabic{table}}

\section*{Appendix B: GINI INDEX OF DECOUPLED DUAL LOG-NORMAL MIXTURE}

For a single log-normal distribution ${\rm Lognormal}(\mu,\sigma^2)$, the Gini index $g$ is given by
\begin{align}
    g={\rm erf}(\sigma/2),
    \label{eq-g-LN}
\end{align}
which is also captured by the Lorenz curve $\mathcal{L}$ that is the cumulative sum of income fraction from the poorest to the richest sample. 
The $\mathcal{L}(f)$ divides the lower triangle into two regions. If upper and lower regions are denoted as $A$ and $B$, respectively, $g$ is given by
\begin{align}
    g\equiv\frac{A}{A+B}=1-2B,
    \label{eq-g-area}
\end{align}
where $A+B=1/2$ and $B=\int_{0}^{1}\mathcal{L}(f)df$. 

When the log-income distribution is a mixture of two distributions, $\rho(x)=[\rho_1+\rho_2]/2$, and is completely separable, \textit{i.e.}, i.e., support sets, $X_1~\mbox{and}~X_2$, satisfy $X_2\ge X_1$ ($h=1$), the entire Lorenz curves $\mathcal{L}$ represented by $\mathcal{L}_1$ and $\mathcal{L}_2$, are rescaled Lorenz curves of $\rho_1$ and $\rho_2$, respectively. 

\begin{figure}[]
\includegraphics[width=\columnwidth]{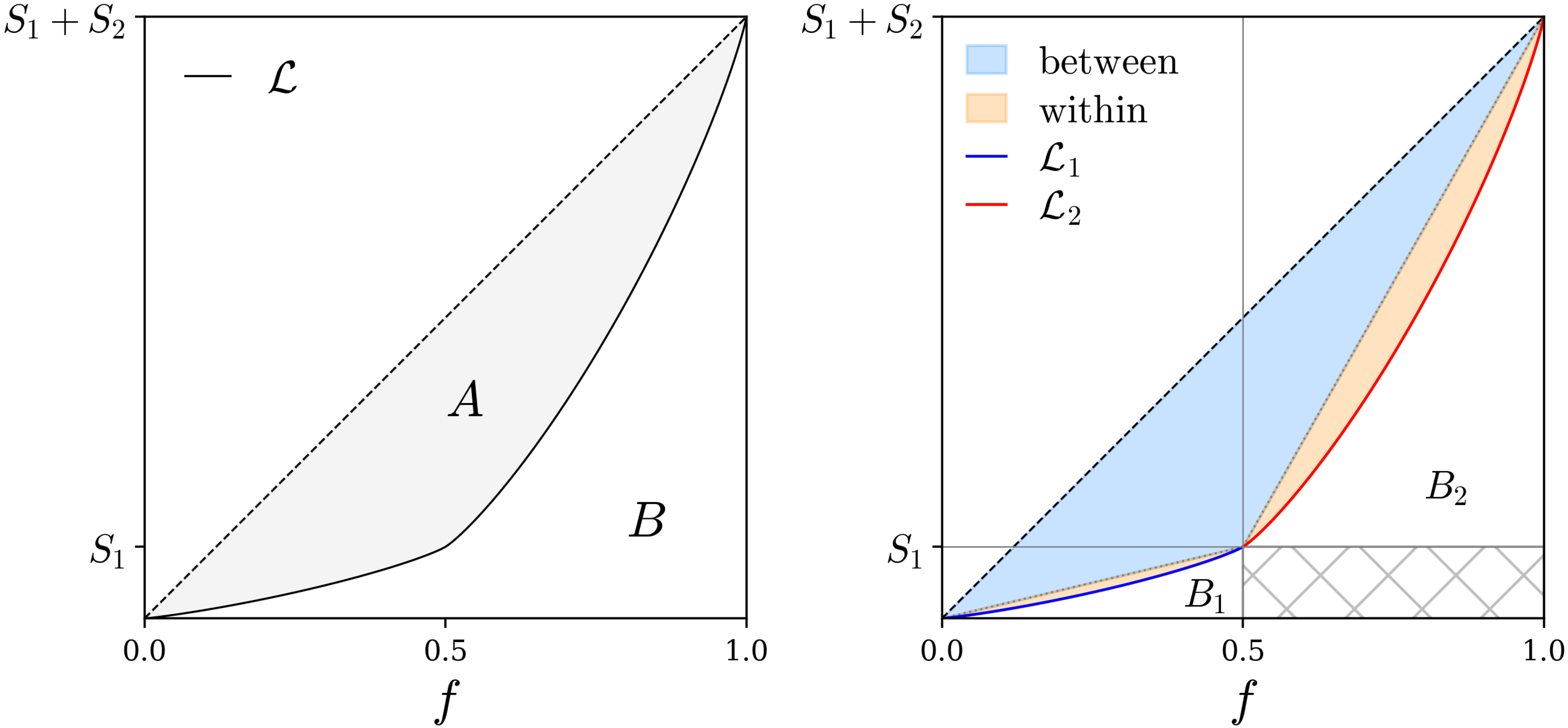}
    \caption{Visualization of Lorenz curve and decomposition for the case of a dual log-normal mixture with large decoupling: Entire Lorenz curve $\mathcal{L}$ (left); Decomposition of $\mathcal{L}$ (right), where blue and red solid lines show rescaled Lorenz curves $\mathcal{L}_1$ and $\mathcal{L}_2$, and $S_1$ and $S_2$ represent total income share from each log-normal distribution. Light-blue and light-orange shadowed areas show contributions of between- and within-inequality on $g$ for $f_1=f_2=1/2$.}
    \label{fig-EM-b1}
\end{figure}

If the total population (income) share of the first and second distributions is given by $f_1$ and $f_2$ ($S_1$ and $S_2$), 
then $B$ is decomposed as [see Fig.~\ref{fig-EM-b1}]
\begin{align}
    B=B_1+B_2+(1-f_1)S_1.
    \label{eq-B}
\end{align}
Since the Lorenz curves, $\mathcal{L}_1$ and $\mathcal{L}_2$, for sufficiently large decoupling ($h\approx1$) correspond to the rescaling of the single Lorenz curve of each distribution, the rescaled areas, $B_1$ and $B_2$, are rewritten as
\begin{align}
    B_1=f_1S_1\frac{1}{2}(1-g_1),
    \quad B_2=f_2S_2\frac{1}{2}(1-g_2),
    \label{eq-B12}
\end{align}
where $g_1$ and $g_2$ are the corresponding Gini indices for $\rho_1$ and $\rho_2$, respectively. Substituting $f_1=f_2=1/2$, Eq.~\eqref{eq-B12} and Eq.~\eqref{eq-B} into Eq.~\eqref{eq-g-area}, $g$ is as
\begin{align}
    g=\frac{1}{2}\left(1+S_1g_1+S_2g_2\right)-S_1.
    \label{g-large-h}
\end{align}
This representation is valid for arbitrary $\rho_1$ and $\rho_2$ with $f_1=f_2=1/2$ and $X_2\ge X_1$. 

For the dual log-normal mixture of ${\rm Lognormal}(\mu_1,\sigma^2)$ and ${\rm Lognormal}(\mu_2,\sigma^2)$, where $\mu_1<\mu_2$, $h\approx1$, and the same fractions $f_1=f_2=1/2$ (population shares for $\rho_1$ and $\rho_2$), the income shares $S_1~\mbox{and}~S_2$ are approximately as 
\begin{align}
    S_1 &= \frac{\langle{c_1}\rangle}{\langle{c_1}\rangle+\langle{c_2}\rangle}=\frac{1}{1+e^{\Delta\mu}}, \nonumber \\ 
    S_2 &= \frac{\langle{c_2}\rangle}{\langle{c_1}\rangle+\langle{c_2}\rangle}=\frac{1}{1+e^{-\Delta\mu}},
    \label{eq-S12}
\end{align}
where $\langle{c_1}\rangle=\exp(\mu_1+\sigma^2/2)$, $\langle{c_2}\rangle=\exp(\mu_2+\sigma^2/2)$ from the log-normal nature, and $\Delta\mu=(\mu_2-\mu_1)$.  
Moreover, for the perfectly assortative case with $\mathcal{A}_{\max}$, $\mu_1=\mu_t^-,\mu_2=\mu_t^+$, $\sigma^2=\sigma_t^2=\beta^2t^\lambda/(2Ja_0)$ and $\Delta\mu=2\Delta\alpha t$ for large $t$.

Substituting $f_1=f_2=1/2$, Eq.~\eqref{eq-S12}, and Eq.~\eqref{eq-g-LN} into Eq.~\eqref{g-large-h}, $g(t)$ becomes
\begin{align}
    g(t)=\frac{1}{2}\left(1-\frac{2}{1+e^{-2\Delta\alpha t}}\right)+\frac{1}{2}{\rm erf}(\sigma_t/2)
    \label{eq-g-Amax}
\end{align}
which corresponds to Eq.~\eqref{gini-Amax}. For small decoupling $h\ll1$, $g$ of the dual log-normal mixture becomes more complicated (see Sec.~III~A in SM~\cite{SM}).
\end{appendix}

\begin{widetext}
\setcounter{equation}{0}
\setcounter{figure}{0}
\setcounter{table}{0}
\setcounter{page}{1}
\makeatletter
\renewcommand{\theequation}{S\arabic{equation}}
\renewcommand{\thefigure}{S\arabic{figure}}
\renewcommand{\thetable}{S\arabic{table}}
\renewcommand{\bibnumfmt}[1]{[#1]}
\renewcommand{\citenumfont}[1]{#1}

\section*{\bf Supplemental Material for\\ ``Anomaly, Class Division, and Decoupling in Income Dynamics''} 

\section{BM model on ring topologies}

This section is the extended version of \textit{Appendix A} in \textbf{End Matter} (EM), where we provide all the details (analytical derivations and numerical confirmations) for the homogeneous Bouchaud-M{\'e}zard (BM) model on ring topologies. 

\subsection{\label{var-cov-eta} Variance, Covariance, and Field exponent $\eta$} 

For a 1D ring topology, Eq.~(A2)
in EM can be rewritten as follows:
\begin{align}
    dc_n=J\left(\frac{c_{n-1}+c_{n+1}}{2}-c_n\right)dt+\beta c_ndW_{t,n},
    \label{eq-c}
\end{align}
where $n=0,1,\dots,N-1$. In the periodic boundary conditions, $c_{-1}=c_{N-1}$ and $c_{N}=c_0$. Let $X_n=\ln{c_n}+\beta^2t/2$ and apply the It{\^o}'s lemma into Eq.~\eqref{eq-c}, it becomes
\begin{align}
    dX_n=d\ln{c_n}+\frac{1}{2}\beta^2dt=J\left(\frac{c_{n-1}+c_{n+1}}{2c_n}-1\right)dt+\beta dW_{t,n}-\frac{1}{2}\beta^2 dt+\frac{1}{2}\beta^2dt.
    \label{eq-X0}
\end{align}
Using $c_n=\exp{(X_n-\beta^2t/2)}$, the equation for $X_n$ can be rewritten as follows: 
\begin{align}
    dX_n=J\left(\frac{1}{2}e^{X_{n-1}-X_n}+\frac{1}{2}e^{X_{n+1}-X_n}-1\right)dt+\beta dW_{t,n}.
    \label{eq-X1}
\end{align}
We assume that for a small $\beta$ condition, the difference of $X$ between neighboring nodes becomes sufficiently small. As a result, the first-order approximation should be:
\begin{align}
    dX_n=J\left(\frac{X_{n-1}+X_{n+1}}{2}-X_n\right)dt+\beta dW_{t,n}.
    \label{eq-X}
\end{align}

To solve the stochastic differential equations (SDE) of Eq.~\eqref{eq-X}, we perform a discrete Fourier transform:
\begin{align}
    \hat{X}_{\rm k}=\frac{1}{\sqrt{N}}\sum_{k=0}^{N-1}X_ne^{-i2\pi\frac{\rm k}{N}n},\quad
    X_n=\frac{1}{\sqrt{N}}\sum_{n=0}^{N-1}\hat{X}_{\rm k}e^{i2\pi\frac{\rm k}{N}n}.
\end{align}
Substituting the discrete inverse Fourier transform into Eq.~\eqref{eq-X}, it becomes
\begin{align}
    d\hat{X}_{\rm k}(t)=-J\left(1-\cos{\frac{2\pi{\rm k}}{N}}\right)\hat{X}_{\rm k}(t)dt+\beta dW_{\rm k}(t),
\end{align}
which is the SDE corresponding to the ${\rm k}$-th Fourier mode. Here we can see that $\hat{X}_{\rm k}(t)$ obeys an independent {\it Ornstein-Uhlenbeck} (OU) process at the mode $k$. The solution of the OU process is
\begin{align}
    \hat{X}_{\rm k}(t)=\hat{X}_{\rm k}(0)e^{-Jt(1-\cos{2\pi {\rm k}/N})}+\beta\int_{0}^{t}e^{-J(1-\cos{2\pi {\rm k}/N})(t-s)}dW_{\rm k}(s)
    \label{OU-sol}
\end{align}
According to the Parseval's theorem, $\sum_{n=0}^{N-1}|X_n|^2=\sum_{k=0}^{N-1}|\hat{X}_{\rm k}|^2$, the variance of $X_n$ is as follows:
\begin{align}
    {\rm Var}(X_n)=\frac{1}{N}\sum_{n=0}^{N-1}\mathbb{E}[|X_n|^2]-\left(\frac{1}{N}\sum_{n=0}^{N-1}\mathbb{E}[X_n]\right)^2=\frac{1}{N}\sum_{{\rm k}=0}^{N-1}\mathbb{E}[|\hat{X}_{\rm k}|^2]-\left(\frac{1}{N}\sum_{n=0}^{N-1}\mathbb{E}[X_n]\right)^2,
    \label{Var-Xn}
\end{align}
where $\mathbb{E}[\cdot]$ is the expectation for the stochastic process. We consider the initial condition $X_n(0)={\rm const}$, where all nodes have the same value of $X$. From the given SDE, $d\left(\sum_{n=0}^{N-1}X_n\right)=0$, so that the mean value of $X_n$ is always a constant, regardless of $t$. For this case, $\hat{X}_{\rm k}(0)=\frac{1}{\sqrt{N}}\sum_{n=0}^{N-1}X_n(0)e^{-i2\pi\frac{\rm k}{N}n}=\sqrt{N}X_n(0)\delta_{0{\rm k}}$. 

To get ${\rm Var}(X_n)$, we calculate $\mathbb{E}[|\hat{X}_{\rm k}|^2]$: 
\begin{align}
    \mathbb{E}[|\hat{X}_{\rm k}|^2] &= \hat{X}_{\rm k}^2(0)e^{-2Jt(1-\cos{2\pi{\rm k}/N})}
    +2\beta\hat{X}_{\rm k}(0)e^{-Jt(1-\cos{2\pi{\rm k}/N})}\int_{0}^{t}e^{-J(1-\cos{2\pi{\rm k}/N})(t-s)}\langle{dW_{\rm k}(s)}\rangle \nonumber \\
    & +\beta^2\int_{0}^{t}\int_{0}^{t}e^{-2J(1-\cos{2\pi{\rm k}/N})(t-\frac{s+s'}{2})}\langle{dW_{\rm k}(s)dW_{\rm k}(s')}\rangle \nonumber \\
    & =\hat{X}_{\rm k}^2(0)e^{-2Jt(1-\cos{2\pi{\rm k}/N})}
    +\beta^2\int_{0}^{t}e^{-2J(1-\cos{2\pi{\rm k}/N})(t-s)}ds \nonumber \\
    & =NX^2_n(0)\delta_{0{\rm k}}e^{-2Jt(1-\cos{2\pi{\rm k}/N})}
    +\beta^2\left[\frac{1-e^{-2Jt(1-\cos{2\pi{\rm k}/N})}}{2J(1-\cos{2\pi{\rm k}/N)}}\right],
\end{align}
where we use $\langle{dW_{\rm k}(s)}\rangle=0$, $\langle{dW_{\rm k}(s)dW_{\rm k}(s')}\rangle=\delta(s-s')ds$, and $\delta_{0{\rm k}}^2=\delta_{0{\rm k}}$.
As a result, we get
\begin{align}
    {\rm Var}(X_n) &= \left[\frac{1}{N}\sum_{{\rm k}=0}^{N-1}NX_n(0)^2\delta_{0{\rm k}}e^{-2Jt(1-\cos 2\pi{\rm k}/N)}
    +\frac{1}{N}\sum_{{\rm k}=0}^{N-1}\beta^2\left[\frac{1-e^{-2Jt(1-\cos{2\pi{\rm k}/N})}}{2J(1-\cos{2\pi{\rm k}/N)}}\right]\right]
    -\left[X_n(0)^2\right] \nonumber \\
    & =\frac{1}{N}\sum_{{\rm k}=0}^{N-1}\beta^2\left[\frac{1-e^{-2Jt(1-\cos{2\pi{\rm k}/N})}}{2J(1-\cos{2\pi{\rm k}/N)}}\right] \approx\frac{\beta^2}{4\pi J}\int_{0}^{2\pi}\frac{1-e^{-2Jt(1-\cos{u})}}{1-\cos{u}}du
    =\frac{\beta^2}{2\pi J}\int_{0}^{\pi}\frac{1-e^{-2Jt(1-\cos{u})}}{1-\cos{u}}du,
\end{align}
which is the result of transforming the discrete sum of Fourier mode ${\rm k}$ into a definite integral for $N\to\infty$, and the periodic property of the integrand allows us to reduce the integration interval $[0,2\pi]$ to $[0,\pi]$. If we differentiate the expression of ${\rm Var}(X_n)$ with respect to $t$,
\begin{align}
    \frac{d}{dt}{\rm Var}(X_n)=\beta^2 e^{-2Jt}\cdot\frac{1}{\pi}\int_{0}^{\pi}e^{2Jt\cos{u}}du=\beta^2e^{-2Jt}I_0(2Jt),
    \label{var-diffeq}
\end{align}
where $I_\ell(z)$ is the modified Bessel function of the first kind, represented by
\begin{align}
    I_\ell(z)=\frac{1}{\pi}\int_{0}^{\pi}e^{z\cos{\theta}}\cos{\ell\theta}d\theta-\frac{\sin{\ell\pi}}{\pi}\int_{0}^{\infty}e^{-z\cosh{q}-\ell q}dq
    \label{Bessel}.
\end{align}

We get ${\rm Var}(X_n)$ by solving Eq.~\eqref{var-diffeq}. The right-hand side function has a well-defined indefinite integral:
\begin{align}
    \int e^{-2Jt}I_0(2Jt)dt=te^{-2Jt}\left[I_0(2Jt)+I_1(2Jt)\right]+{\rm C}.
\end{align}
According to initial condition $X_n(0)={\rm const}$ at $t=0$, ${\rm Var}(X_n(0))=0$ and  $\lim_{t\to0}te^{-2Jt}\left[I_0(2Jt)+I_1(2Jt)\right]=0$, and a constant of integration ${\rm C}=0$. Therefore, 
\begin{align}
    {\rm Var}(X_n)=\beta^2te^{-2Jt}[I_0(2Jt)+I_1(2Jt)].
    \label{eq-var-X}
\end{align}
The Taylor expansions of Eq.~\eqref{eq-var-X} for small $t$ and large $t$, respectively, are 
\begin{align}
    \begin{cases}
    \beta^2t\left(1-\frac{(2Jt)}{2}+\frac{(2Jt)^2}{4}-\frac{5(2Jt)^3}{48}+\cdots\right) & \text{for small } t, \\
    \beta^2t\left(\sqrt{\frac{2}{\pi}}(2Jt)^{-1/2}-\frac{1}{4\sqrt{2\pi}}(2Jt)^{-3/2}-\frac{3}{64\sqrt{2\pi}}(2Jt)^{-5/2}-\cdots\right) &\text{for large } t,
    \end{cases}
\end{align}
where the leading order terms are $\beta^2t$ and $\beta^2\sqrt{t}/(2Ja_0)$, respectively.

With $a_0(J)=\sqrt{\pi/(4J)}$, we finally get
\begin{align}
    {\rm Var}(X_n)=
    \begin{cases}
    \beta^2 t & {\rm for\ small\ }t,\\
    \beta^2\sqrt{t}/(2Ja_0) & {\rm for\ large\ }t.
    \end{cases}
\end{align}
Thus, a given stochastic process in Eq.~\eqref{eq-X} starts with normal diffusion and ends with sub-diffusion. For large $t$, the analytical results of ${\rm Var}(X_n)\sim\sqrt{t}$ is consistent with the EFT and numerical simulations.

\begin{figure*}[]
    \includegraphics[width=\textwidth]{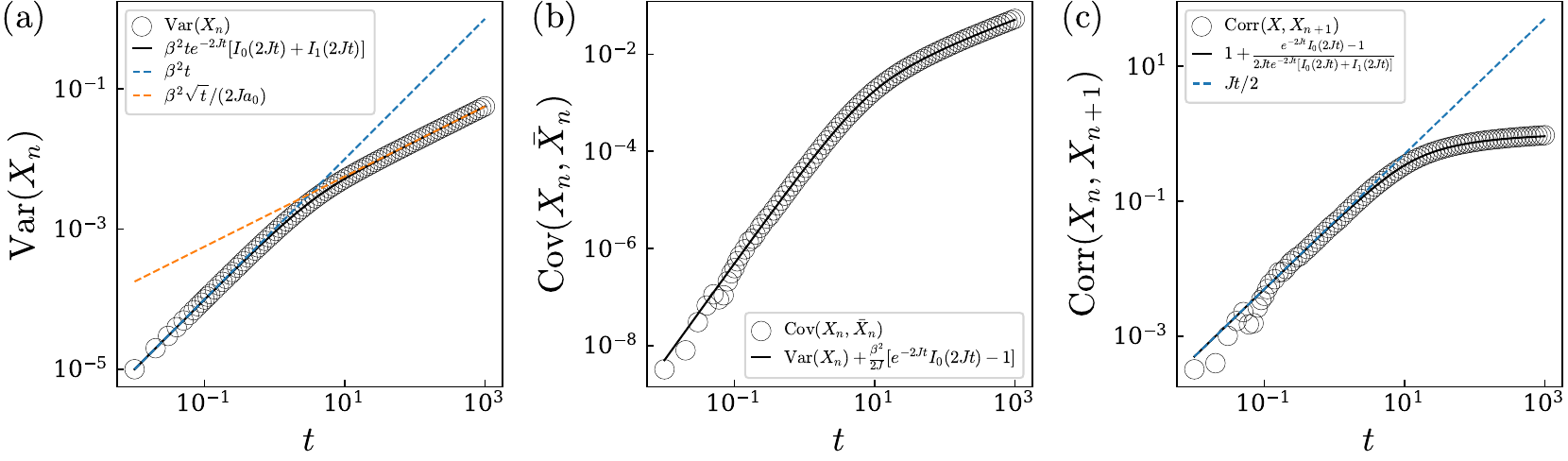}
    \caption{(a) Variance ${\rm Var}(X_n)$, (b) covariance ${\rm Cov}(X_n,\bar{X}_n)$, and (c) correlation ${\rm Corr}(X_n,X_{n+1})$ in the BM model for a 1D ring topology. Open symbols ($\circ$) represent numerical simulation results and and solid lines represent the theoretical prediction. In (a), the blue dashed line $\{\beta^2t,Jt/2\}$ is the predictions for small $t$ and othe range dashed line $\{\beta^2\sqrt{t}/(2Ja_0)\}$ is the prediction for large $t$. For all cases, $N=10^4,\alpha=10^{-2},\beta^2=10^{-3},J=10^{-1}$, and all data are averaged over 128 ensembles.}
    \label{vcc}
\end{figure*}
To obtain the field exponent $\eta$ related to the diffusion nature, we require the calculation of ${\rm Cov}(X_n,\bar{X}_n)$. First, the discrete Fourier transform of $\bar{X}_n$ is:
\begin{align}
    \hat{\bar{X}}_{\rm k} &= \frac{1}{\sqrt{N}}\sum_{n=0}^{N-1}\bar{X}_ne^{-i2\pi\frac{\rm k}{N}n} = \frac{1}{\sqrt{N}}\sum_{n=0}^{N-1}\frac{X_{n-1}+X_{n+1}}{2}e^{-i2\pi\frac{\rm k}{N}n} \nonumber \\
    & =\frac{1}{2\sqrt{N}}\sum_{n=0}^{N-1}\left[\frac{1}{\sqrt{N}}\sum_{{\rm k}'=0}^{N-1}(\hat{X}_{{\rm k}'}e^{i2\pi\frac{{\rm k}'}{N}(n-1)}+\hat{X}_{{\rm k}'}e^{i2\pi\frac{{\rm k}'}{N}(n+1)})\right]e^{-i2\pi\frac{\rm k}{N}n} \nonumber \\
    & =\frac{1}{2\sqrt{N}}\sum_{n=0}^{N-1}\left[\frac{1}{\sqrt{N}}\sum_{{\rm k}'=0}^{
    N-1}\hat{X}_{{\rm k}'}e^{i2\pi\frac{{\rm k}'}{N}n}(e^{-i2\pi\frac{{\rm k}'}{N}}+e^{i2\pi\frac{{\rm k}'}{N}})\right]e^{-i2\pi\frac{\rm k}{N}n} \nonumber \\
    & =\frac{1}{\sqrt{N}}\sum_{n=0}^{N-1}\left[\frac{1}{\sqrt{N}}\sum_{{\rm k}'=0}^{N-1}\cos{\left(\frac{2\pi{\rm k}'}{N}\right)}\hat{X}_{{\rm k}'}e^{i2\pi\frac{{\rm k}'}{N}n}\right]e^{-i2\pi\frac{\rm k}{N}n} =\frac{1}{N}\sum_{{\rm k}'=0}^{N-1}\cos{\left(\frac{2\pi{\rm k}'}{N}\right)}\hat{X}_{{\rm k}'}\left[\sum_{n=0}^{N-1}e^{-i2\pi(\frac{{\rm k}-{\rm k}'}{N})n}\right] \nonumber \\
    & =\frac{1}{N}\sum_{{\rm k}'=0}^{N-1}\cos{\left(\frac{2\pi{\rm k}'}{N}\right)}\hat{X}_{{\rm k}'}N\delta_{{\rm kk}'} =\cos{\left(\frac{2\pi{\rm k}}{N}\right)}\hat{X}_{\rm k}.
\end{align}
Substituting this result into the Parseval's theorem, we get
    $\sum_{n=0}^{N-1}X_n\bar{X}_n^*
    =\sum_{k=0}^{N-1}\hat{X}_{\rm k}\hat{\bar{X}}_{\rm k}^*
    =\sum_{k=0}^{N-1}\cos{\left(\frac{2\pi{\rm k}}{N}\right)|\hat{X}_{\rm k}|^2}$,
where $*$ denotes complex conjugate. So the covariance between $X_n$ and $\bar{X}_n$,
${\rm Cov}(X_n,\bar{X}_n)$ is as follows:
\begin{align}
    {\rm Cov}(X_n,\bar{X}_n) &= \frac{1}{N}\sum_{n=0}^{N-1}\mathbb{E}[X_n\bar{X}_n^*]-\left(\frac{1}{N}\sum_{n=0}^{N-1}\mathbb{E}[X_n]\right)\left(\frac{1}{N}\sum_{n=0}^{N-1}\mathbb{E}[\bar{X}_n]\right) \nonumber \\
    & = \frac{1}{N}\sum_{k=0}^{N-1}\cos{\left(\frac{2\pi{\rm k}}{N}\right)}\mathbb{E}[|\hat{X}_{\rm k}|^2]-\left(\frac{1}{N}\sum_{n=0}^{N-1}\mathbb{E}[X_n]\right)\left(\frac{1}{N}\sum_{n=0}^{N-1}\mathbb{E}[\bar{X}_n]\right).
\end{align}
By repeating the same procedure for the derivation of ${\rm Var}(X_n)$, we get
\begin{align}
    {\rm Cov}(X_n,\bar{X}_n)=\frac{\beta^2}{2\pi J}\int_{0}^{\pi}\cos{u}\left[\frac{1-e^{-2Jt(1-\cos{u})}}{1-\cos{u}}\right]du, 
\end{align}
and its time derivative is:
\begin{align}
    \frac{d}{dt}{\rm Cov}(X_n,\bar{X}_n)=\beta^2e^{-2Jt}I_1(2Jt).
\end{align}

The indefinite integral is:
\begin{align}
    \int d[{\rm Cov}(X_n,\bar{X}_n)] &= \beta^2\int e^{-2Jt}I_1(2Jt)dt
    = \beta^2\int e^{-2Jt}(1/2J)\left[\frac{d}{dt}I_0(2Jt)\right]dt \nonumber \\
    & =\frac{\beta^2}{2J}\left[e^{-2Jt}I_0(2Jt)+2J\int e^{-2Jt}I_0(2Jt)dt\right]
    =\frac{\beta^2}{2J}e^{-2Jt}I_0(2Jt)+{\rm Var}(X_n)+{\rm C}.
\end{align}
Since the initial condition ${\rm Cov}(X_n(0),\bar{X}_n(0))=0$ and $\lim_{t\to 0}e^{-2Jt}I_0(2Jt)=1$ give ${\rm C}=-\frac{\beta^2}{2J}$. Therefore, the covariance becomes
\begin{align}
    {\rm Cov}(X_n,\bar{X}_n)={\rm Var}(X_n)+\frac{\beta^2}{2J}\left[e^{-2Jt}I_0(2Jt)-1\right].
\end{align}
The correlation between $X_n$ and $X_{n+1}$,
${\rm Corr}(X_n,X_{n+1})$ is as follows:
\begin{align}
    {\rm Corr}(X_n,X_{n+1})\equiv\frac{{\rm Cov}(X_n,X_{n+1})}{\sqrt{{\rm Var}(X_n){\rm Var}(X_{n+1})}}=\frac{{\rm Cov}(X_n,\bar{X}_n)}{{\rm Var}(X_n)}.
\end{align}
where ${\rm Cov}(X_n,\bar{X}_n)=(1/2)\left[{\rm Cov}(X_n,X_{n-1})+{\rm Cov}(X_n,X_{n+1})\right]={\rm Cov}(X_n,X_{n+1})$ and ${\rm Var}(X_n)={\rm Var}(X_{n+1})$ due to the translational invariance of the system. For small $t$, $c\simeq 1+x$, so that ${\rm Var}(X_n)\simeq{\rm Var}(c_n)$ and ${\rm Cov}(X_n,\bar{X}_n)\simeq{\rm Cov}(c,\bar{c}_n)$. We note that the linear increase of variance and correlation for small $t$ was also revealed by Medo~\cite{medo2009breakdown}. Our theory for the BM model in a 1D ring not only supports the earlier result but also predicts variance, covariance, and correlation for large $t$.

\begin{figure*}[]
    \includegraphics[width=\textwidth]{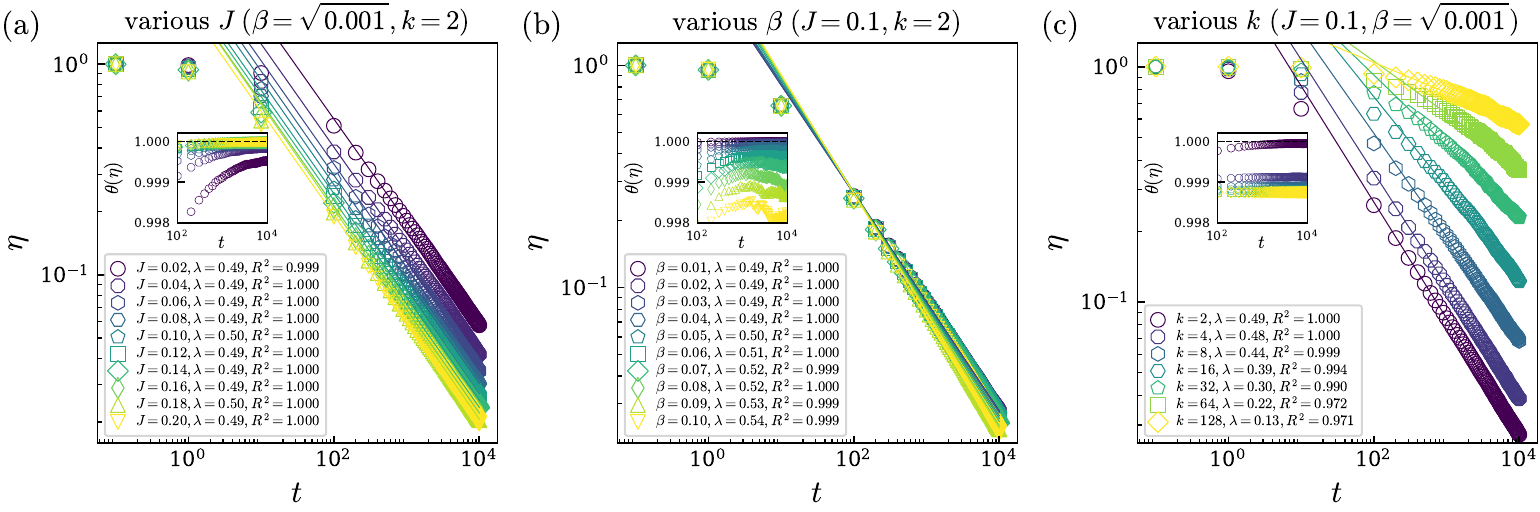}
    \caption{The field exponent $\eta(t)$ for a variety of the condition  ($J,\beta,k$). (a) The case with various $J$ and sufficiently small $\beta=\sqrt{0.001}$. (b) The case various $\beta$. (c) The case with various $k$, where $k$ is the number of neighbors in a regular network of 1D ring topology. Open symbols ($\circ$) represent numerical simulation results, and solid lines represent the least square linear regression of data samples for $t\geq 10^2$. $R^2$ represents the coefficient of determination for regression lines. Insets show $\theta(\eta)$ for $t\geq10^2$. For all cases, $N=10^4$, all data are averaged over 128 ensembles.}
    \label{eta}
\end{figure*}
From the results of ${\rm Var}(X_n)$ and ${\rm Cov}(X_n,\bar{X}_n)$, we get 
\begin{align}
    1-\eta=\frac{{\rm Cov}(X_n,\bar{X}_n)}{{\rm Var}(X_n)} = \frac{{\rm Var}(X_n)+\frac{\beta^2}{2J}\left[e^{-2Jt}I_0(2Jt)-1\right]}{{\rm Var}(X_n)}=1+\frac{e^{-2Jt}I_0(2Jt)-1}{2Jte^{-2Jt}\left[I_0(2Jt)+I_1(2Jt)\right]}.
\end{align}
As a result, we can calculate the field exponent $\eta(t)$:
\begin{align}
    \eta(t)=\frac{1-e^{-2Jt}I_0(2Jt)}{2Jte^{-2Jt}\left[I_0(2Jt)+I_1(2Jt)\right]}.
    \label{eq-eta}
\end{align}
Since $\eta(t)=-\frac{1}{2J}\frac{I_0(2Jt)}{t[I_0(2Jt)+I_1(2Jt)]}+\frac{\beta^2}{2J}\frac{1}{{\rm Var}(X_n)}$ and $\lim_{t\to\infty}\frac{I_0(2Jt)}{t[I_0(2Jt)+I_1(2Jt)]}=0$, for large $t$, $\eta(t)=\frac{\beta^2}{2J}\frac{2Ja_0}{\beta^2}t^{-1/2}=a_0t^{-1/2}$ which matches our numerical simulation results for small $\beta$ condition.

Figure~\ref{eta} shows the time evolution of $\eta$ for a variety of the condition $(J,\beta,k)$. For small $\beta$, Eq.~\eqref{eq-eta} tells us that the power-law decaying exponent $\lambda=1/2$ is independent of $J$. Since $a_0(J)=\sqrt{\pi/(4J)}$, an increase in $J$ is only attributed to the effect of parallel shifting $\eta$ in double-logarithmic plots. However, for large $\beta$, $\lambda>1/2$. For this case, the difference in $X$ between neighbors cannot be approximated to the first order because it is more affected by multiplicative noise in Eq.~\eqref{eq-c}. Thus, the variance grows faster than the order of $\sqrt{t}$. The number of neighbors $k$ also can change the value of $\eta(t)$. For sufficiently small $k$, $\eta(t)$ still exhibits a power-law decay with $\lambda<1/2$, which is no longer valid for large $k$. This was also reported in study by Ma et al.~\cite{ma2013distribution} as follows: If $k$ is large enough, $\eta$ has a non-zero finite value and $\rho_{\rm eq}(c)$ satisfying the Fokker-Planck equation is the generalized inverse-gamma distribution. As a result, $\eta(t)\sim a_0t^{-\lambda}$ for sufficiently small $\beta$ and $k$. Thus, the given SDE in Eq.~\eqref{eq-c} and Eq.~(A3) of \textit{Appendix A} in EM of the main text can be approximated by the time-dependent {\it Ornstein-Uhlenbeck} process through the EFT ansatz as in Eq.~(A4) of \textit{Appendix A} in EM of the main text for such conditions. Therefore,
\begin{align}
    {\rm Var}(X_n)=
    \begin{cases}
        \beta^2t &\text{for small $t$},\\
    \beta^2/(2J\eta)=\beta^2t^\lambda/(2Ja_0) &\text{for large $t$},
    \end{cases}
    \label{Var-general}
\end{align}
where $1/2\leq\lambda\leq1$ for the 1D case. For small $\beta$, we observe that $\lambda=1/2$ in Eq.~\eqref{eq-eta}, and for large $\beta$, the interaction term becomes relatively small, compared to the fluctuation term, which makes the system behave closer to normal diffusion, so that $\lambda$ does not exceed 1 (the value for the ballistic behavior).

We conclude this subsection (\ref{var-cov-eta}) with a discussion on the mean of $x$. In fact, deterministic drifts to all $X_n$ only give a translation to the probability density function but do not change its shape. Thus, ${\rm Var}(X_n)$ is independent of any homogeneous drift term, so that we arbitrarily define $X_n=\ln c_n+\beta^2t/2$ in Eq.~\eqref{eq-X}. Since the normalized income is defined as $c\equiv C/\bar{C}$, it satisfies
$\langle{c}\rangle=1$. In addition, $x$ follows a log-normal distribution, its mean is $\exp(\mu_t+\sigma_t^2/2)$, and the corresponding drift is $\mu_t=-\sigma_t^2/2$ as we mentioned in the main text as well as EM of the main text. Therefore, the drift of $x$ is not derived by the underlying equation in Eq.~\eqref{eq-X}, but rather by the normalization drift that $c$ is rescaled at every step.

\subsection{\label{LRC} 
Long-range correlation analysis}

The covariance ${\rm Cov}(X_n,\bar{X}_n)$ in the previous subsection (\ref{var-cov-eta}) is the log-income correlation between neighboring nodes. We here go further and consider correlations between nodes located at greater distances. For any node $n$, let $\bar{X}_{(n,\ell)}$ be the average log-income of nodes of distance $\ell$, then $\bar{X}_{(n,\ell)}=(X_{n-\ell}+X_{n+\ell})/2$. Using the same process in \ref{var-cov-eta}, we get
\begin{align}
    \frac{d}{dt}{\rm Cov}(X_n,\bar{X}_{(n,\ell)})=\beta^2e^{-2Jt}I_{\ell}(2Jt).
    \label{eq-diff-cov}
\end{align}
The function $e^{-2Jt}I_{\ell}(2Jt)$ on the right-hand side has a leading order of $(2Jt)^{\ell}$ for small $t$. The correlation between nodes separated by a distance $\ell$ is as follows:
\begin{align}
    {\rm Corr}(X_n,X_{n+\ell})=\frac{{\rm Cov}(X_n,X_{n+\ell})}{\sqrt{{\rm Var}(X_n){\rm Var}(X_{n+\ell})}} = \frac{{\rm Cov}(X_n,\bar{X}_{(n,\ell)})}{{\rm Var}(X_n)}.
    \label{l-ranged-corr}
\end{align}
We note that ${\rm Var}(X_n)={\rm Var}(X_{n+\ell})$ by the translational invariance and be careful for ${\rm Cov}(X_n,X_{n+\ell})={\rm Cov}(X_n,\bar{X}_{(n,\ell)})$ but ${\rm Corr}(X_n,X_{n+\ell})\neq{\rm Corr}(X_n,\bar{X}_{(n,\ell)})$. For small $t$, ${\rm Cov}(X_n,\bar{X}_{(n,\ell)})\sim t^{\ell+1}$ and a leading term of ${\rm Var}(X_n)\sim t$, so that the correlation becomes
\begin{align}
    {\rm Corr}(X_n,X_{n+\ell})\sim t^\ell,\quad{\rm for\ }t\ll1.
    \label{eq-corr-order-early}
\end{align}
Since the initial condition $c_n(0)=1$ at $t=0$, the first-order approximation is valid for small $t$. Thus, ${\rm Corr}(c_n,c_{n+\ell})\simeq{\rm Corr}(1+x_n,1+\bar{x}_{n+\ell})={\rm Corr}(x_n,\bar{x}_{n+\ell})$. Therefore, we get 
\begin{align}
    {\rm Corr}(c_n,c_{n+\ell})\sim t^\ell,\quad{\rm for\ }t\ll1.
\end{align}
This result theoretically supports the correlation analysis in the early time, which was reported by Medo~\cite{medo2009breakdown}. Moreover, we can show the expression of ${\rm Corr}(X_n,X_{n+\ell})$ for large $t$. As a result, Eq.~\eqref{eq-X} is rewritten as a multi-dimensional OU process:
\begin{align}
    d{\bf X}_t={\mathbb K}{\bf X}_tdt+\beta d{\bf W}_t,
\end{align}
where ${\bf X}_t=(X_0,\dots,X_{N-1}),{\bf W}_t=(W_{t,0},\dots,W_{t,N-1})$ and ${\mathbb K}$ is a $N\times N$ stability matrix:
\begin{align}
    {\mathbb K}=J
    \begin{bmatrix}
        -1 & 1/2 & 0 & \cdots & 1/2 \\
        1/2 & -1 & 1/2 & \cdots & 0 \\
        0 & 1/2 & -1 & \cdots & 0 \\
        \vdots & \vdots & \vdots & \ddots & \vdots \\
        1/2 & 0 & 0 & \cdots & -1
    \end{bmatrix},
\end{align}
\begin{figure}[]
    \includegraphics[width=\textwidth]{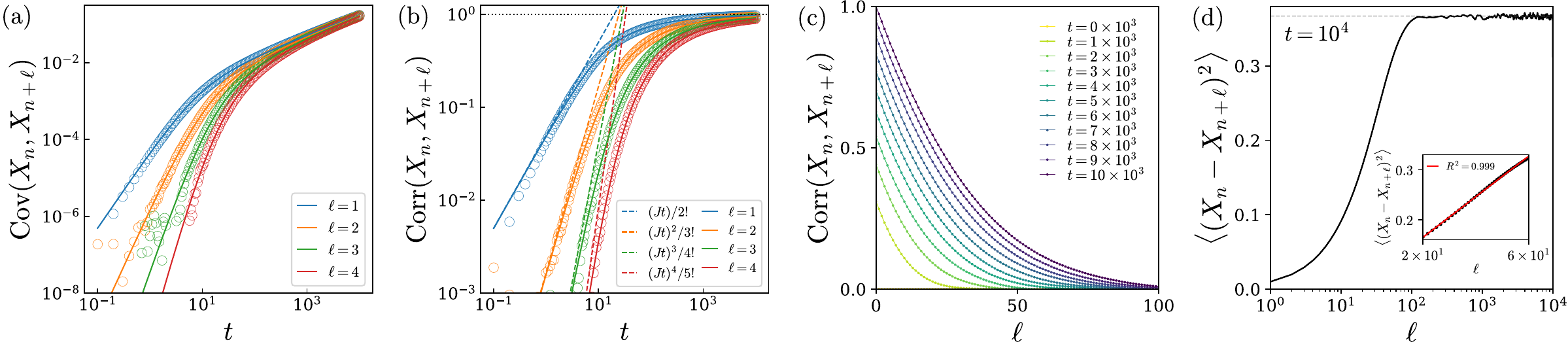}
    \caption{(a) ${\rm Cov}(X_n,X_{n+\ell})$ and (b) ${\rm Corr}(X_n,X_{n+\ell})$ against  time $t$ for $\ell=\{1,2,3,4\}$. For (a) and (b), open symbols ($\circ$) are numerical simulation results, solid lines are theoretical predictions, and dashed lines are the leading order of the Taylor expansion in Eq.~\eqref{eq-covl1234} at $t=0$. (c) ${\rm Corr}(X_n,X_{n+\ell})$ for $\ell\in[0,100]$ and (d) $\langle{(X_n-X_{n+\ell})^2}\rangle$ for $\ell\in[0,10^4]$. The inset of (d) shows $\langle{(X_n-X_{n+\ell})^2}\rangle$ for $\ell\in[20,60]$. For (c) and (d), solid lines are numerical simulation results. For all cases, $N=10^4,\alpha=10^{-2},\Delta\alpha=10^{-3},\beta^2=10^{-3},J=10^{-1}$, and all data are averaged over 128 ensembles.}
    \label{l-range}
\end{figure}
The stability of the system is determined by the eigenvalues of ${\mathbb K}$. There are three cases: (1) Stable case, ${\rm Re}(\lambda_n)<0$ for all $n$; (2) Marginally stable case, ${\rm Re}(\lambda_n)\leq0$ and at least one $\lambda_n=0$; (3) Unstable case, ${\rm Re}(\lambda_n)>0$ for at least one $n$. Since ${\mathbb K}$ is circulant matrix, eigenvalues are as follows: 
\begin{align}
    \lambda_n=-J\left(1-\cos{\frac{2\pi n}{N}}\right)\leq0 \quad(n=0,1,\dots,N-1).
\end{align}
Thus, ${\mathbb K}$ is marginally stable~\cite{bouchaud2024self}. For this case, the system shows a strongly auto-correlated behavior for long-time scales. The $\ell$-ranged covariance can be obtained by a definite integral of Eq.~\eqref{eq-diff-cov}. Denoting $Q_{\ell}=\beta^2\int_{0}^{t}e^{-2Js}I_{\ell}(2Js)ds$, a covariance matrix ${\bf\Sigma}_t$ is as follows:
\begin{align}
    {\bf\Sigma}_t=\beta^2
    \begin{bmatrix}
        Q_{0} & Q_{1} & \cdots & Q_{N/2-1} & Q_{N/2} & Q_{N/2+1} & \cdots & Q_{1} \\
        Q_{1} & Q_{0} & \cdots & Q_{N/2-2} & Q_{N/2-1} & Q_{N/2} & \cdots & Q_{2} \\
        \vdots & \vdots & & \vdots & \vdots & \vdots & & \vdots \\
        Q_{1} & Q_{2} & \cdots & Q_{N/2} & Q_{N/2-1} & Q_{N/2-2} & \cdots & Q_{0} \\
    \end{bmatrix},
\end{align}
where the expression for $N$ is even. For the same time $t$, we check that ${\rm Cov}(X_n,X_{n+\ell})$ decreases against $\ell$. The analytical expressions of the covariance for $\ell=\{1,2,3,4\}$ are as follows:
\begin{equation}
    {\rm Cov}(X_n,X_{n+\ell})=
    \begin{cases}
        \frac{\beta^2}{2J}\left[-1+(1+2Jt)e^{-2Jt}I_0(2Jt)+2Jte^{-2Jt}I_1(2Jt)\right] & (\ell=1)\\
        \frac{\beta^2}{2J}\left[-2+(2+2Jt)e^{-2Jt}[I_0(2Jt)+I_1(2Jt)]\right] & (\ell=2) \\
        \frac{\beta^2}{2J}\frac{1}{2Jt}\left[-3(2Jt)+2Jt(5+2Jt)e^{-2Jt}I_0(2Jt)+[-4+2Jt(4+2Jt)]e^{-2Jt}I_1(2Jt)\right] & (\ell=3)\\
        \frac{\beta^2}{2J}\frac{1}{(2Jt)^2}\left[-4(2Jt)^2+16e^{-2Jt}I_1(2Jt)+2Jt[-8+2Jt(8+2Jt)]e^{-2Jt}[I_0(2Jt)+I_1(2Jt)]\right] & (\ell=4)
    \end{cases}
    \label{eq-covl1234}
\end{equation}
${\rm Corr}(X_n,X_{n+\ell})$ in Eq.~\eqref{eq-covl1234} has a leading order as $\sim(Jt)^\ell/(\ell+1)!$ in the early time. Moreover, Eq.~\eqref{eq-covl1234} matches for large $t$, so that we can investigate the case for larger distances $\ell$.
The long-range spatial correlation could be shown by the spatial variogram $\mathcal{V}(\ell)\equiv\langle{(X_n-X_{n+\ell})^2}\rangle$. Since $\mathcal{V}(\ell)$ increases, in proportional to $\sim\log\ell$, until it saturates for large $\ell$ [see the inset of Fig.~\ref{l-range}~(d)], the 1D BM model results in the long-range spatial correlation with large clusters in poor and rich classes, compared to other models [see Fig.~\ref{ACC-models}~(a)].

\begin{figure}[]
    \includegraphics[width=\textwidth]{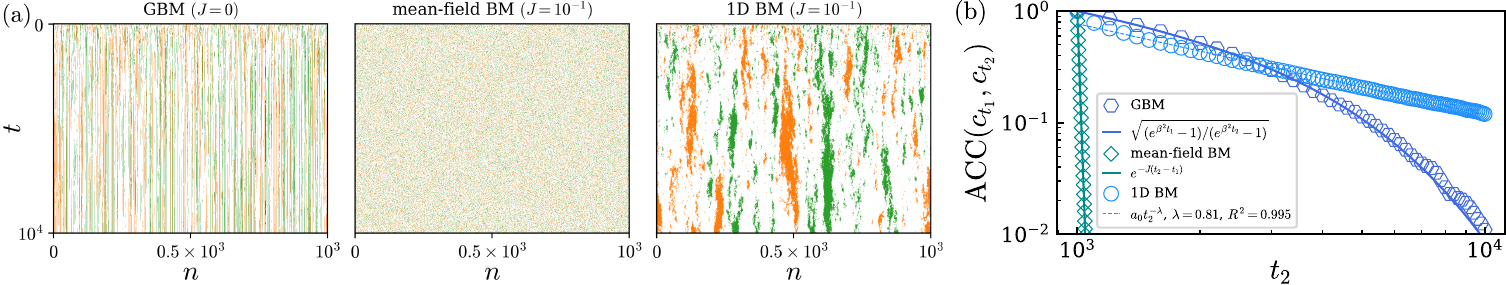}
    \caption{(a) Spatiotemporal patterns of top-rich/bottom-poor 10\% (orange/green) class in homogeneous growth-rate models: the geometric Brownian motion (GBM) in the left panel, the mean-field BM model in the middle panel, and the 1D BM model in the right panel. (b) ${\rm ACC}(c_{t_1},c_{t_2})$ for the three models. Open symbols ($\circ$) represent numerical simulation results, solid lines represent theoretical predictions, and the dashed line is the regression line. Here $t_1=10^3$. For all cases, $\alpha=10^{-2}$ and $\beta^2=10^{-3}$. ACCs are obtained by $N=10^4$ with 128 ensembles.}
    \label{ACC-models}
\end{figure}

\subsection{\label{ACC} Autocorrelation coefficient for various network topologies}

We investigate the persistence of income in the BM model by the autocorrelation coefficient~\cite{park2018fundamentals} (ACC), and compare it with other two versions: (1) the geometric Brownian motion (GBM, non-interactive case) and (2) the mean-field BM model (fully-connected case). The ACC is defined as:
\begin{align}
    {\rm ACC}(c_{t_1},c_{t_2})\equiv\frac{{\rm Cov}(c_{t_1},c_{t_2})}{\sqrt{{\rm Var}(c_{t_1}){\rm Var}(c_{t_2})}}.
\end{align}
In the GBM, by using 
$$C_t=C(0)\exp{\left[\left(\alpha-\beta^2/2\right)t+\beta W_t\right]} 
\mbox{~~and~~}
\mathbb{E}[e^{X+Y}]=\exp{\left[\mu_X+\mu_Y+(\mu_X^2+\mu_Y^2)/2+{\rm Cov}(X,Y)\right]}$$ 
for dependent Gaussian random variables $X$ and $Y$, we get the following statistics: (1) ${\rm Var}(c_t)=e^{2\alpha t}(e^{\beta^2t}-1)$, and (2) ${\rm Cov}(c_t,c_s)=e^{\alpha(t+s)}\left[e^{\beta^2\min{(t,s)}}-1\right]$.
Thus, ${\rm ACC}(c_t,c_s)=\sqrt{(e^{\beta^2s}-1)/(e^{\beta^2t}-1)}$ where $t>s$ and ${\rm ACC}(c_t,c_s)\approx e^{-\beta^2|t-s|/2}$ for large $(t,s)$. For the mean-field BM model, ACC is known as $e^{-J\Delta t}$ for the condition of $2J/\beta^2>1$~\cite{bouchaud2000wealth}.In contrast to these cases, which show exponential decay of ACC, for the 1D ring topology, ACC exhibits a power-law decay [see Fig.~\ref{ACC-models}~(b)] with long-term persistence of income.

\section{Statistical properties of binary mixture in 1D ring}

In this section, we provide all the details for the statistical properties of binary growth rates, $\alpha_\pm=\alpha\pm \Delta\alpha$, in a 1D ring. Here $\alpha > \Delta\alpha >0$.

\subsection{Assortativity $\mathcal{A}$}
We define the growth rate assortativity as $\mathcal{A}\equiv{\rm Cov}(\alpha,\alpha')/{\sqrt{{\rm Var}(\alpha){\rm Var}(\alpha')}}$, where $\alpha$ and $\alpha'$ are the growth rates of neighboring nodes and $\alpha\in\{\alpha-\Delta\alpha,\alpha+\Delta\alpha\}$. If we let $N_{\pm}=N/2$ be the number of nodes with a lower (higher) growth rate, we simply find that ${\rm Var}(\alpha)={\rm Var}(\alpha')=\frac{1}{N}\left[\frac{N}{2}(\alpha-\Delta\alpha)^2+\frac{N}{2}(\alpha+\Delta\alpha)^2\right]-\alpha^2=\Delta\alpha^2$. For a 1D ring topology, the number of links is the same as the number of nodes $N$. The number of nodes can be decomposed as $N=N_{ll}+N_{hh}+N_{lh}$, where $N_{ll}$ ($N_{hh}$) is the number of links between both lower (higher) $\alpha$ nodes and $N_{lh}$ is the number of links between the lower and higher $\alpha$ nodes. For this case,
\begin{align}
    {\rm Cov}(\alpha,\alpha')
    & =\frac{1}{N}\left[ N_{ll}(\alpha-\Delta\alpha)^2+N_{hh}(\alpha+\Delta\alpha)^2+N_{lh}(\alpha-\Delta\alpha)(\alpha+\Delta\alpha) \right]-\alpha^2 \nonumber \\
    & =\frac{N_{ll}+N_{hh}-N_{lh}}{N}\Delta\alpha^2+(N_{hh}-N_{ll})(2\alpha\Delta\alpha) =\frac{N^{(\rm 1)}-N^{(\rm 2)}}{N}\Delta\alpha^2.
\end{align}
To reflect the changes in the link configuration for each pair of node swapping, we must consider two neighboring values $\alpha$ of the selected nodes. Each node can be categorized by three consecutive network motifs. For this case, there are six motifs and the possible cases of center node swapping between motifs are $\binom{6}{2}$. For all possible cases, $\Delta N_{ll}=\Delta N_{hh}$, so $N_{ll}-N_{hh}$ is conserved under any pair node swapping. Therefore, for the case of $N_-=N_+$, $N_{ll}=N_{hh}$. If we let homogeneous and heterogeneous link densities be $\rho^{(1)}=(N_{ll}+N_{hh})/N$ and $\rho^{(2)}=N_{lh}/N$, then $\mathcal{A}=\rho^{(1)}-\rho^{(2)}$ for the case of $N_-=N_+$.

\begin{figure}[]
    \includegraphics[width=\textwidth]{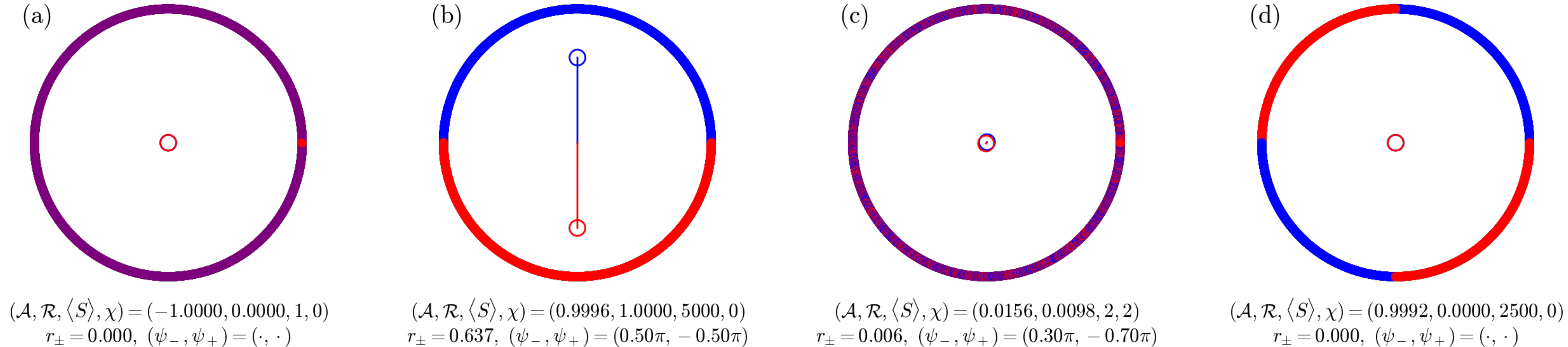}
    \caption{Statistical properties $(\mathcal{A},\mathcal{R},\langle{S}\rangle,\chi)$ and order parameters $(r,\psi)$ for a binary mixture in ring topologies: (a) Perfectly disassortative configuration $\mathcal{A}_{\min}$. (b) Perfectly assortative configuration $\mathcal{A}_{\max}$. (c) Randomly permuted configuration. (d) Regularly clustered configuration of $\langle{S}\rangle=2.5\times10^3$. For all cases, $N=10^4$ and $N_{\pm}=5\times 10^3$.}
    \label{order-parameters}
\end{figure}
\subsection{Kuramoto oscillator order parameters $(r,\psi)$: Concentration $\mathcal{R}$}

We consider the case of $N_-=N_+$, where the number of each binary element is the same. For a perfectly assortative configuration ($\mathcal{A}_{\max}$) with large $N$, we can calculate $(r_-,\psi_-)$ and $(r_+,\psi_+)$ as follows:
\begin{align}
    r_-e^{i\psi_-}
    &=\frac{1}{N/2}\sum_{j=1}^{N/2}e^{i(2\pi j/N)}
    =\frac{1}{2\pi}\frac{2\pi}{N/2}\sum_{j=1}^{N/2}e^{i(2\pi j/N)};~~
    \frac{1}{\pi}\int_{0}^{\pi}e^{iu}du
    = \frac{2}{\pi}i, \\
    r_+e^{i\psi_+}
    &=\frac{1}{N/2}\sum_{j=N/2+1}^{N}e^{i(2\pi j/N)}
    =\frac{1}{2\pi}\frac{2\pi}{N/2}\sum_{j=N/2+1}^{N}e^{i(2\pi j/N)};~~
    \frac{1}{\pi}\int_{\pi}^{2\pi}e^{iu}du
    = -\frac{2}{\pi}i.
\end{align}
Based on these results, we get $(r_-,\psi_-)=\left(\frac{2}{\pi},\frac{\pi}{2}\right)$ and 
$(r_+,\psi_+)=\left(\frac{2}{\pi},\frac{3\pi}{2}\right)$, where $2/\pi$ is the maximum value of $r_{\pm}$. 
For a perfectly assortative configuration, we show that the following relations are satisfied:
\begin{align}
    r_- &= r_+,\quad \Delta\psi=|\psi_--\psi_+|=\pi.
    \label{r-psi}
\end{align}
The whole configuration can be generated by pair node swapping that starts from a perfectly assortative configuration. Thus, if the relations of Eq.~\eqref{r-psi} are robust under arbitrary pair node swappings, we can address that they are universal properties for the HBM model with a binary mixture of growth rates in the 1D ring.

For an initial binary configuration, 
its order parameters $\vec{r}_-,\vec{r}_+$ are represented by the real vector form. We assume that the initial configuration satisfies $\vec{r}_-=-\vec{r}_+$. When arbitrary nodes $(j,k)$ are swapped, there are two cases: (1) $(j,k)$ are in the same binary element group and (2) $(j,k)$ are in different binary element groups. For the first case, order parameters do not change. For the second case, the order parameters of the swapped configuration are:
\begin{align}
    \vec{r}_- '&= \vec{r}_--\vec{a}_j+\vec{a}_k=\vec{r}_-+\vec{A}, \\
    \vec{r}_+ '&= \vec{r}_++\vec{a}_j-\vec{a}_k=\vec{r}_+-\vec{A},
\end{align}
where $\vec{a}_j=\left(\cos{\frac{2\pi j}{N}},\sin{\frac{2\pi j}{N}}\right)$ and $\vec{A}=\vec{a}_k-\vec{a}_j$. For the case of $\vec{r}_-=-\vec{r}_+$, $\vec{r}_+'=\vec{r}_+-\vec{A}=-(\vec{r}_-+\vec{A})=-\vec{r}_-'$. Thus, $\vec{r}_+'$ has the same length and the opposite direction of $\vec{r}_-'$. Therefore, Eq.~\eqref{r-psi} is satisfied under arbitrary pair node swapping, except the cases of $\vec{r}_-'=0$ or $\vec{r}_+'=0$. \textit{Path 1} and \textit{Path 2} in the main text [see Fig.~1~(d)] satisfies statistical properties, only except for a perfectly disassortative configuration $(\mathcal{A}_{\min},\mathcal{R}_{\min})$. There are several cases that satisfy $r=0$, excluding a perfectly disassortative configuration. If the same binary elements are allocated in exactly opposite directions for all locations, $r=0$. For this case, the angular argument $\psi$ could not be defined.

\subsection{Cluster-size distribution $P(S)$, $\langle S\rangle$, and $\chi$}

The size $S$ distribution of the clusters, $P(S)$, can be characterized by the average $\langle{S}\rangle$ and the variance $\chi$. For the case of $N_-=N_+$ as $N\to\infty$, a binary mixture on a 1D ring is just a random sequence of binary elements with the same probability $p=1/2$. If a cluster is defined as a consecutive sequence of the same binary element, the probability mass function of $S$ is as follows:
\begin{align}
    P(S)=\big(\frac{1}{2}\big)^{S+1},
\end{align}
so that 
$$\langle{S}\rangle=\mathbb{E} [S]=\sum_{S=1}^{\infty}S\cdot P(S)=2~~\mbox{and}~~
\chi=\mathbb{E}[S^2]-\mathbb{E}[S]^2=2$$
for the random configuration as $N\to\infty$. In particular, if all cluster sizes are exactly the same, this configuration satisfies $\mathcal{R}=0$ with $\chi=0$ and $N/(2\langle{S}\rangle)\in\mathbb{N}$. 
For regularly clustered configurations, $\langle{S}\rangle$ affects the spatiotemporal patterns of income dynamics.  

\section{\label{HBM} Heterogeneous BM model in 1D ring}

In this section, we provide all the details (analytical derivations and numerical confirmations) for the heterogeneous BM (HBM) model in a 1D ring, which is compared to our findings in the main text.

\begin{figure}[b]
    \includegraphics[width=\textwidth]{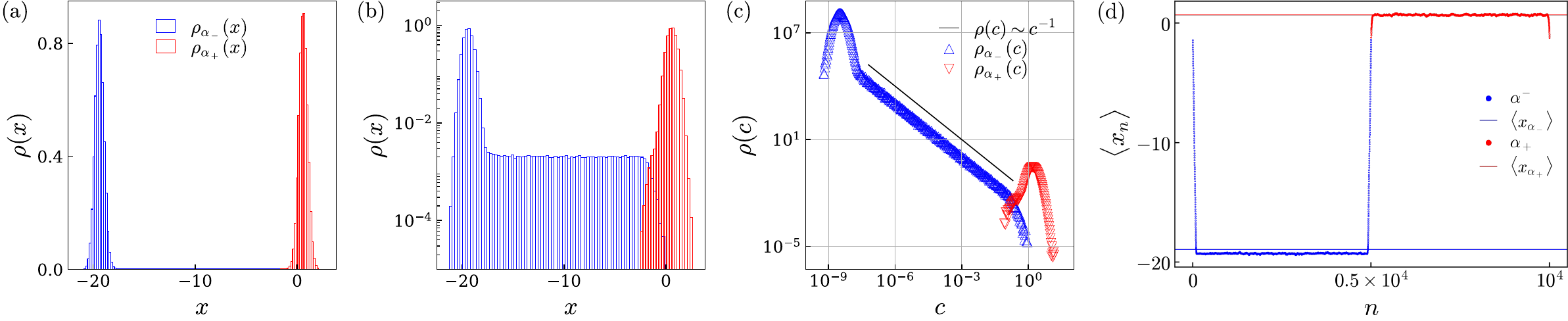}
    \caption{Log-income ($x=\ln c$) distribution for the case of $\mathcal{A}_{\max}$: (a) $\rho_{\alpha_\pm}(x)$ and (b) the semi-logarithmic scaled plot of $\rho_{\alpha_\pm}(x)$. (c) Double-logarithmic scaled plot of $\rho(c)$, where the black solid line represents the guided line of $\rho(c)\sim c^{-1}$. (d) The positional mean of $x$, $\langle{x_n}\rangle$. For all cases, $N=10^4,\alpha=10^{-2},\Delta\alpha=10^{-3},\beta^2=10^{-3},J=10^{-1},t=10^4$ and all data are averaged over 128 ensembles.}
    \label{Amax}
\end{figure}
\subsection{Normalized income distribution $\rho(x)$: Perfectly disassortative \textit{versus} Perfectly assortative}

Let $\rho(x),\rho_{\alpha_-}(x)$, and $\rho_{\alpha_+}(x)$ be the probability density functions of $x$ for the entire, the lower and higher $\alpha$ groups, respectively. 
For a perfectly disassortative growth rate configuration with $\mathcal{A}_{\min}$, $\rho_{\alpha_-}(x)$ and $\rho_{\alpha_+}(x)$ overlap almost perfectly each other, and the overall distribution is almost the same as that in the BM model. Therefore, $\rho(c,t;\mathcal{A}_{\min})\sim{\rm Lognormal}(\mu_t,\sigma_t^2)$.
For a perfectly assortative growth rate configuration with $\mathcal{A}_{\max}$, both of $\rho_{\alpha_\pm}(x)$ do not overlap almost perfectly each other. For this case, $\rho(x,t)$ can be roughly divided into three regions: (1) \textit{Head} -- the first peak; (2) \textit{Body} -- the middle part between the first and second peaks; (3) \textit{Tail} -- the second peak. In Fig.~\ref{Amax}~(a)-(c), we observe more precisely how many samples exist between the two peaks. Surprisingly, we find that the samples of $x$ between the two peaks follow a uniform distribution, where most of them belong to the $\alpha_-$ group. It implies that for the case of $\mathcal{A}_{\max}$, $\rho_{\alpha_-}(x)$ and $\rho_{\alpha_+}(x)$ are asymmetric over $x$, whereas for the case of $\mathcal{A}_{\min}$, they are symmetric. In Fig.~\ref{Amax}~(c), the double logarithmic scaled plots show that the body region follows a power-law of $\rho(c)\sim A c^{-1}$, where $A$ is a constant. This is a natural consequence of the distribution transformation formula: $\rho(c)=\rho(x)|dx/dc|=A|d(\ln{c})/dc|=Ac^{-1}$. Our hypothesis in the regional separation of the distribution function is that the field exponent $\eta$ depends on the positional index of a node $n$.

Equation~(A6) in \textit{Appendix A} of EM, is obtained by the first-order approximation with the assumption that $\eta x$ is sufficiently small. However, if $\eta$ is extremely small, the interaction term $J\left[\theta(\eta)c^{1-n}-c\right]dt$ in Eq.~(A5) can be neglected due to $c^{1-\eta}\to c$ and $\theta(\eta)\to 1$. Therefore, Eq.~(A5) becomes separated into two cases as follows:
\begin{align}
    dc_n=J[\theta(\eta)c_n^{1-\eta}-c_n]dt+\beta c_ndW_{t,n},
    \nonumber\\    
    ~\longrightarrow
    \begin{cases}
    dx_n &= J\eta_{t,n}[\mu_{t,n}-x_n]dt+\beta dW_{t,n} \\
    dc_n &= \beta c_ndW_{t,n}
    \end{cases}
    \label{eq-cases}
\end{align}
If our hypothesis is correct, the governing equation for $\rho(c,t)$ should depend on the node's position $n$ for the given network. In Fig.~\ref{Amax}~(d), we plot $\langle{x_n}\rangle$ against $n$, which confirms our hypothesis. In the near body region between two different $\alpha$ clusters, $\langle{x_n}\rangle$ is almost equidistantly spaced and proportional to $n$. Since $\langle{x_n}\rangle$ is obtained by the ensemble average at position $n$, there is the same number of node samples for each value. Therefore, $x$ drawn from these samples of nodes becomes uniformly distributed in $\rho(x)$. Since the number of samples is the same at each point, $\rho^{(\rm b)}(x,t)={\rm const}$. Therefore, $\frac{\partial}{\partial t}\rho^{(\rm b)}(c,t)=0$. 

The Fokker-Planck equation for the second case of Eq.~\eqref{eq-cases} is as follows:
\begin{align}
    \frac{\partial}{\partial t}\rho(c,t)=\frac{1}{2}\frac{\partial^2}{\partial c^2}\left[\beta^2c^2\rho(c,t)\right].
\end{align}
For the body region, $\frac{\partial}{\partial t}\rho^{(\rm b)}(c,t)=0$ and its solution is $\rho^{(\rm b)}(c,t)=Ac^{-1}+Bc^{-2}$, where $A$ and $B$ are constants. By numerical simulations, we observe $\rho^{(\rm b)}(c,t)\sim Ac^{-1}$, which supports $B=0$. Thus, the governing equation for $c_n$ in this region is equal to the second case of Eq.~\eqref{eq-cases}, and $\eta$ measured from these samples is extremely small. For this case, we call the set of nodes the body class if $c$ follows power-law as $\rho(c)\sim Ac^{-1}$. Automatically, the regions excluding the body class should be the head class, represented by small $\langle{x_n}\rangle$, and the tail class, represented by large $\langle{x_n}\rangle$, respectively [see Fig.~\ref{Amax}~(d)].
Head and tail classes exhibit power-law decays, which are the same as those in the BM model [see Fig.~\ref{SDM}~(e) for the field exponent $\eta(t)$ in each region], while $\eta$ for the body class decreases much faster. At $t=10^4$, the order of $\eta$ is smaller than $10^{-4}$. As a result, the interaction term in Eq.~\eqref{eq-cases} can be neglected.
The dynamics of head and tail classes is governed by the first case of Eq.~\eqref{eq-cases}, described by the log-normal distribution, while the dynamics of the body class is governed by the second case of Eq.~\eqref{eq-cases}, which follows a power-law distribution [see Fig.~\ref{Amax}~(c)].
The remainder is the drift $\mu_{t,n}$ for head and tail classes. To figure it out, we consider the null model, where the two $\alpha$ groups are completely separated by two 1D rings. This implies that there is no body region. For this case, each group can be treated as an independent BM model, and the underlying equation is the same as Eq.~(1) in the main text. The average income $C$ of them becomes $\langle{C_{\alpha_{\pm}}(t)}\rangle=C(0)e^{\alpha_{\pm}t}$. Then, the corresponding average normalized income $c$ for each group is as follows:
\begin{align}
    \langle{c_{t,\alpha_\pm}}\rangle_{(\rm null)}= \frac{C(0)e^{(\alpha\pm\Delta\alpha)t}}{[C(0)e^{(\alpha-\Delta\alpha)t}+C(0)e^{(\alpha+\Delta\alpha)t}]/2} = \frac{2}{1+e^{\mp2\Delta\alpha t}},
\end{align}
Since each group follows a log-normal distribution that has the variance $\sigma_t^2=\beta^2t^{\lambda}/(2Ja_0)$ for large $t$, similar to the homogeneous BM model, the normalization conditions are $\langle{c_{t,\alpha_{\pm}}}\rangle_{(\rm null)}=\exp\left(\mu_t^{\pm}+\sigma_t^2/2\right)$. Therefore,
\begin{align}
    \mu_t^{\pm} \equiv\langle{x_{t,\alpha_\pm}}\rangle_{(\rm null)} = -\sigma_t^2/2+\ln\left(\frac{2}{1+e^{\mp 2\Delta\alpha t}}\right),~~ 
    \Delta\mu \equiv|\mu_t^+-\mu_t^-| 
    \approx\begin{cases}
        0 &\text{for small $t$}, \\
        2\Delta\alpha t &\text{for large $t$}.
    \end{cases}
    \label{segregation-null}
\end{align}
As a result, the interval between two Gaussian peaks is linearly proportional to $t$ for large $t$. The larger the difference of two growth rates ($\Delta\alpha$), the wider the segregation of two income levels ($\Delta\mu$).
\begin{figure}[]
    \includegraphics[width=\textwidth]{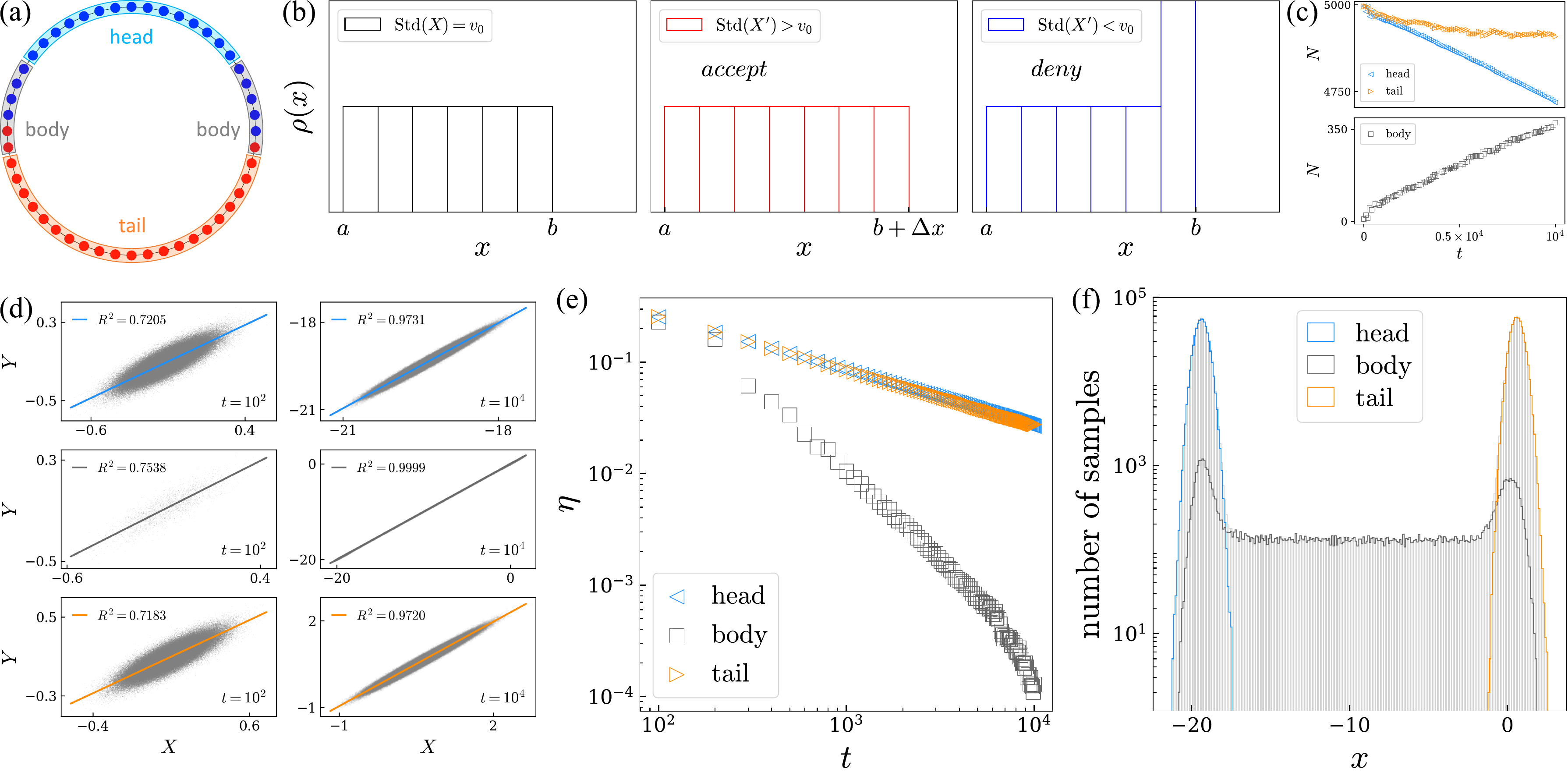}
    \caption{Positional separation for the case of $\mathcal{A}_{\max}$: (a) Conceptual visualization of three parts for the separation of governing equations in a 1D ring. (b) Standard deviation maximization algorithm. Detecting the body region, we start from an active node (the node that has a heterogeneous link) and select a set of nodes within the range $[d_1,d_2]$ from the active node. The standard deviation of $X$, ${\rm Std}(X)$, from theses node samples is maximized in the range $[d_1,d_2]$. (c) The number of nodes belonging to each region, $\mathcal{N}$ against time $t$ for three regions: [top] tail (orange) and head (blue); [bottom] body (black). (d) Correlation between $X=\ln{c}$ and $Y=\ln{\bar{c}_n}$ for three regions at $t=10^2$ and $t=10^4$: Head (top), body (middle), and tail (bottom). (e) Field exponent $\eta(t)$ for head (blue), body (black), and tail regions (orange) against $t$. (f) Portion of $x$ samples for head, body, and tail regions. In (c)-(f), $N=10^4,\alpha=10^{-2},\Delta\alpha=10^{-3},\beta^2=10^{-3},J=10^{-1}$, and all data are averaged over 128 ensembles.}
    \label{SDM}
\end{figure}

We empirically find that $\langle{c_t^{(\rm h)}}\rangle,\langle{c_t^{(\rm t)}}\rangle,\langle{x_t^{(\rm h)}}\rangle,~\mbox{and}~\langle{x_t^{(\rm t)}}\rangle$ are the same as those of the null model [see Fig.~\ref{segregation}~(a)-(d)]. In addition, the segregation of income levels of tail and head nodes, $\Delta\mu'\equiv|\mu_t^{(\rm t)}-\mu_t^{(\rm h)}|$, is compared to $\Delta\mu'-\Delta\mu\sim-b_0t$, where $b_0\ll\Delta\alpha$ [see Fig.~\ref{segregation}~(e) and the inset]. As a result, $\mu_t^{(\rm h)}\approx\mu_t^-$ and $\mu_t^{(\rm t)}\approx\mu_t^+$. Most of the nodes in the uniform distribution region belong to the $\alpha_-$ group and only a small fraction belongs to the $\alpha_+$ group. As the distance between two Gaussian peaks increases as time elapses, the uniform distribution area becomes broader because samples that belong to the Gaussian peaks are absorbed into the uniform distribution region over time. At this moment, in the $\alpha_-$ group, more samples are absorbed than in the $\alpha_+$ group. Therefore, the relative height difference of two Gaussian peaks grows as the uniform distribution region widens over time. This is due to a completely finite $N$ effect. Figure~\ref{SDM}~(c) shows the number of samples for each class, which are summarized as follows [see Eq.~(7) in the main text]:
\begin{align}
    \rho^{(\rm h)}(c,t) \sim {\rm Lognormal}\left(\mu_t^-,\sigma_t^2\right);~ 
    \rho^{(\rm b)}(c,t) \sim Ac^{-1};~ 
    \rho^{(\rm t)}(c,t) \sim {\rm Lognormal}\left(\mu_t^+,\sigma_t^2\right). 
    \label{dist-Amax}
\end{align}
Therefore, we conclude that if growth rates are perfectly segregated in a 1D periodic lattice (ring), the log-income distribution is represented by a mixture of log-normal and power-law distributions, and the income level segregation between two growth rate ($\alpha=\alpha_{\pm}$) groups increases as $\Delta\mu\approx2\Delta\alpha t$. We note that the difference in $\eta$ changes the effective governing equation for $c_n$ [see Eq.~\eqref{eq-cases}], resulting in two different types of distributions.

\subsection{Gini index $g$} 

For the case of $\mathcal{A}_{\max}$, the contribution of body samples to the Gini index $g$ becomes negligible since the number of body samples is very small [see Fig.~\ref{Amax}~(a)]. If a probability distribution $\rho(c)$ represents the dual log-normal mixture of ${\rm Lognormal}(\mu_t^-,\sigma_t^2)$ and ${\rm Lognormal}(\mu_t^+,\sigma_t^2)$ with the fraction of $f_1~\mbox{and}~f_2$, $g$ is as follows:
\begin{align}
    g=\frac{f_1^2e^{\mu_t^-}{\rm erf}(\frac{\sigma_t}{2})+f_2^2e^{\mu_t^+}{\rm erf}(\frac{\sigma_t}{2})+f_1f_2\left[e^{\mu_t^-}{\rm erf}\left(\frac{\mu_t^--\mu_t^+-\sigma_t^2}{2\sigma_t}\right)+e^{\mu_t^+}{\rm erf}\left(\frac{\mu_t^+-\mu_t^--\sigma_t^2}{2\sigma_t}\right)\right]}{f_1e^{\mu_t^-}+f_2e^{\mu_t^+}}.
\end{align}
For our case, $f_1=f_2=1/2$, $\mu_t^+>\mu_t^-$, and $(\mu_t^+-\mu_t^-)\gg\sigma_t^2$ for large $t$. Thus, $g\approx\frac{1}{2}\left(1-\frac{2}{1+e^{2\Delta\alpha t}}\right)+\frac{1}{2}{\rm erf}\left(\frac{\sigma_t}{2}\right)$ for large $t$, which is consistent with Eq.~(9) in the main text and Eq.~(B7) in \textit{Appendix B} [see Fig.~3~(b) in the main text].

\begin{figure}[]
    \includegraphics[width=\textwidth]{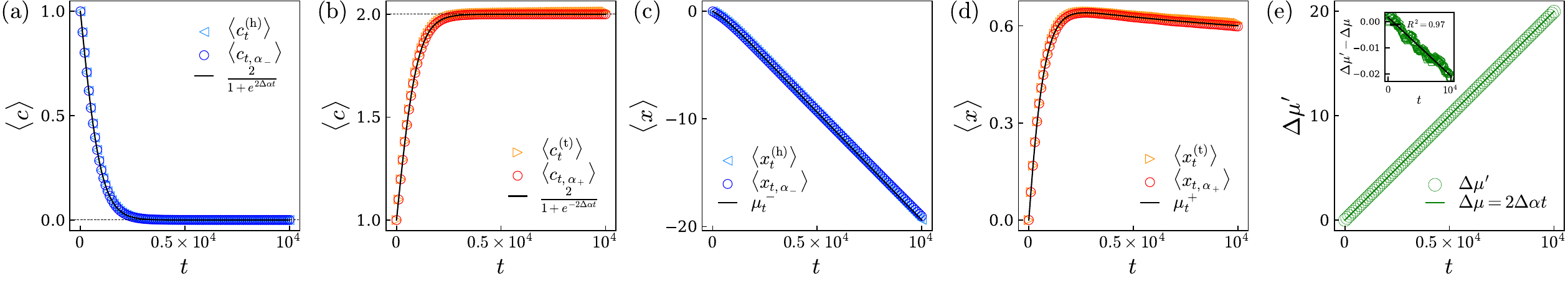}
    \caption{Ensemble averages and income level segregation for $\mathcal{A}_{\max}$ case. (a), (b) Time evolutions of $\langle{c}\rangle$ for head-tail classes, and $\alpha_{\pm}$ groups. (c), (d) Time evolutions of $\langle{x}\rangle$ for head-tial classes and $\alpha_{\pm}$ groups. (e) Income level segregation between two Gaussian peaks $\Delta\mu'$ (see Fig.~\ref{SDM}-(f)). The inset shows $\Delta\mu'-\Delta\mu$. For all cases, $N=10^4,\alpha=10^{-2},\Delta\alpha=10^{-3},\beta^2=10^{-3},J=10^{-1}$ and results are obtained by 128 ensembles.}
    \label{segregation}
\end{figure}

\subsection{Income level segregation $\Delta \mu$}

For the case of $\mathcal{A}_{\max}$, the income level segregation of the $\alpha_{\pm}$ groups is $\Delta\mu\approx2\Delta\alpha t$. However, in general, $\Delta\mu$ depends on the $\alpha$ configurations. For $\mathcal{R}\sim0$ ({\it Path 1}), $\rho_{\alpha_\pm}$ overlap almost each other, so that $\Delta\mu\sim0$ [see Fig.~\ref{x-segregation}~(a) and (c)]. For $\mathcal{R}>0$ ({\it Path 2}), the nodes with $\alpha_-$ can be a rich class and vice versa. Thus, $\Delta\mu$ decreases as the overlap of $\rho_{\alpha_\pm}$ increases [see Fig.~\ref{x-segregation}~(b) and (d)]. For those cases, we empirically find that
\begin{align}
    \Delta\mu\sim
    \begin{cases}
        0 & \text{for {\it Path 1}: $(\mathcal{A}\leq0,\mathcal{R}\sim0)$}, \\
        \mathcal{A}\times 2\Delta\alpha t & \text{for {\it Path 2}: $(\mathcal{A}>0,\mathcal{R}>0)$}.
    \end{cases}
    \label{dmu}
\end{align}
Here $\Delta\mu$ has a very small value for $\mathcal{R}\sim0$, which is controlled by the assortativity $\mathcal{A}$ for $\mathcal{R}>0$ [see Fig.~\ref{x-segregation}~(a) and (b)]. We confirm that this linear relationship is almost valid for $t\sim10^4$. However, for very large $t$, the distribution becomes irregular, which cannot represent two peaks. This seems to be a finite $N$ effect since $\rho_{\alpha_\pm}(x,t)$ for large $t$ strongly depends on local disorder in the configuration.

\begin{figure}[b]
    \centering
    \includegraphics[width=\textwidth]{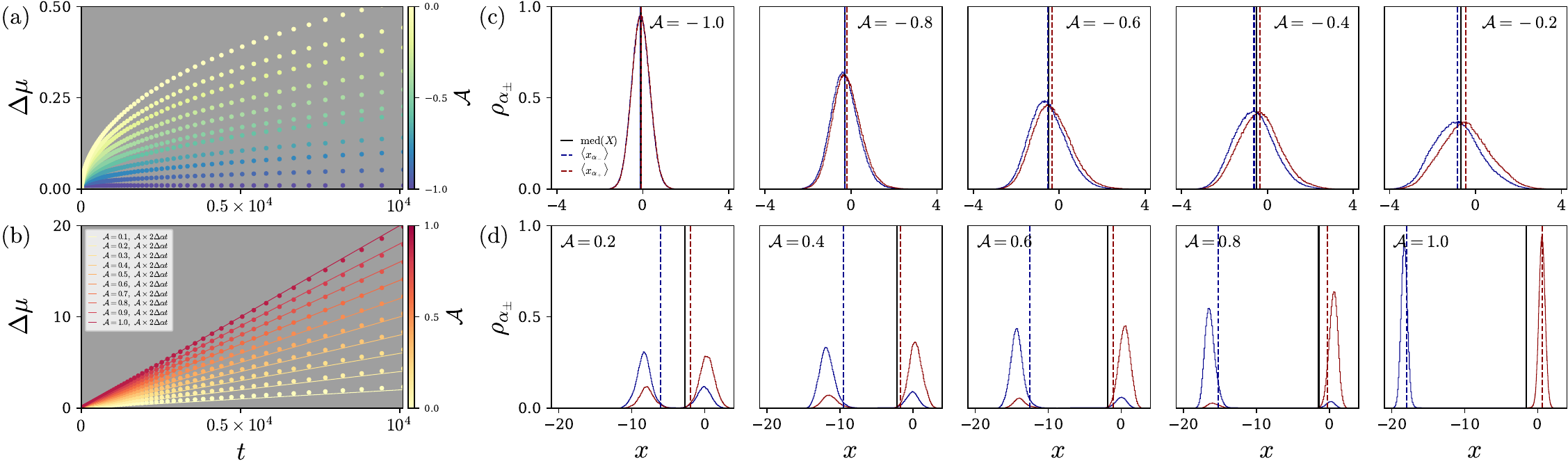}
    \caption{Income level segregation $\Delta\mu$ and distribution $\rho_{\alpha_\pm}(x)$ for various $\alpha$ configurations chosen from random pair swapping, {\it Path 1} and {\it Path 2} [see Fig.~1 in the main text]: For $\Delta\mu$, (a) {\it Path 1} and (b) {\it Path 2}. For $\rho_{\alpha_\pm}(x,t)$ at $t=10^4$, (c) {\it Path 1} and (d) {\it Path 2}. For all cases, $\alpha=10^{-2},\Delta\alpha=10^{-3},\beta^2=10^{-3},J=10^{-1}$ and all data are averaged over 128 ensembles.}
    \label{x-segregation}
\end{figure}

\begin{figure}[]
    \includegraphics[width=\textwidth]{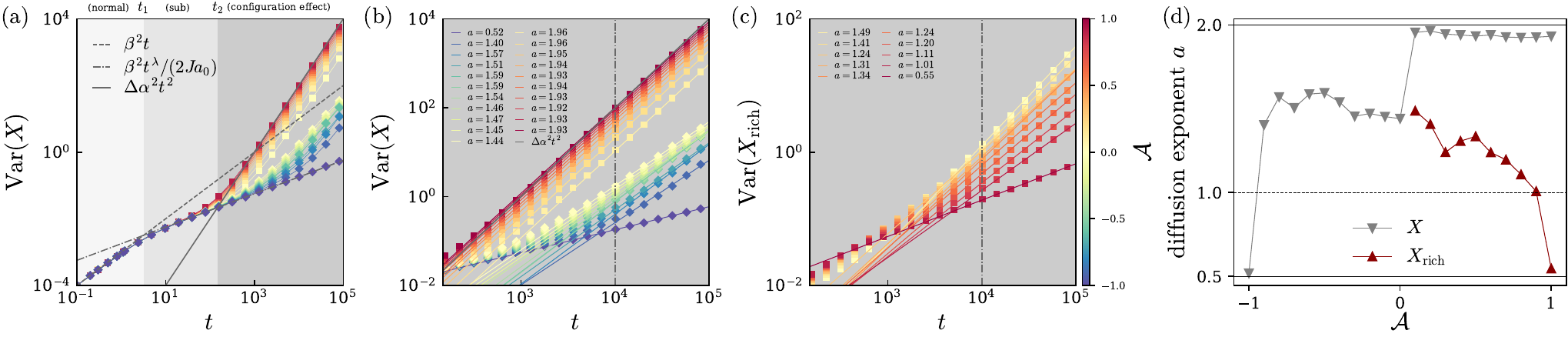}
    \caption{Variances in the HBM model for various $\alpha$ configurations from random pair swapping, {\it Path 1} ($\diamond$) and  {\it Path 2} ($\square$) [see Fig.~1~(d) in the main text]. (a) ${\rm Var}(X)$, (b) ${\rm Var}(X)$ for $t>t_2$, (c) ${\rm Var}(X_{\rm rich})$ for $t>t_2$ and $\mathcal{A}>0$, and (d) diffusion exponent $a$, where $X_{\rm rich}$ is the samples that excess the median of $X$. Colored solid lines show the linear regression for samples with $t>10^4$, and the diffusion exponent $a$ is described in the legends. Here $\alpha=10^{-2}, \Delta\alpha=10^{-3},\beta^2=10^{-3}, J=10^{-1}$, and all data are averaged over 128 ensembles.}
    \label{diffusion-pattern}
\end{figure}

\subsection{Diffusive nature and ballistic motion: Variance ${\rm Var}(X)$}

If the variance of $X$ follows a power law as ${\rm Var}(X)\sim t^{a}$, the diffusion exponent $a$ characterizes the anomaly of diffusive behaviors. In the HBM model for a 1D ring topology, the total variance of $X$ passes through three regimes: (1) Normal diffusion, (2) sub-diffusion, and (3) configuration-effect dominant diffusion. $\rho(x,t;\mathcal{A}_{\max})$ can be approximated as a dual Gaussian mixture of $\mathcal{N}(\mu_t^-,\sigma_t^2)$ and $\mathcal{N}(\mu_t^+,\sigma_t^2)$. As a result,
\begin{align}
    {\rm Var}(X;\mathcal{A}_{\max})\approx\sigma_t^2+\frac{(\mu_t^+-\mu_t^-)^2}{4}
    =\begin{cases}
        \beta^2t &\text{for small $t$},\\
        \beta^2t^{\lambda}/(2Ja_0)+\Delta\alpha^2t^2 &\text{for large $t$},
    \end{cases}
\end{align}
where $|\mu_t^+-\mu_t^-|=\Delta\mu$ in Eq.~\eqref{segregation-null}. Since $0.5\leq\lambda\leq1$, ${\rm Var}(X;\mathcal{A}_{\max})\approx\Delta\alpha^2t^2$ for very large $t$. 

Figure~\ref{diffusion-pattern}~(a) shows that the variance ${\rm Var(X)}$ is characterized by triple time scales: 
$\{\beta^2t,\beta^2t^{\lambda}/(2Ja_0),\Delta\alpha^2t^2\}$. The corresponding intersections are $t_1=[2Ja_0]^{1/(\lambda-1)}$ and $t_2=[2Ja_0\Delta\alpha^2/\beta^2]^{1/(\lambda-2)}$. For $t_1<t<t_2$, the system deviates from the normal diffusion regime of the GBM and enters the sub-diffusion regimes in the BM model. For $t>t_2$, the system deviates from sub-diffusion and enters the configuration-effect dominant regime of the HBM model. This is the unique feature of the HBM model since the BM model always ends with sub-diffusion. From the analytical form of $t_2$, we can easily expect that the sub-diffusion regime vanishes if $\Delta\alpha \gg \beta$.

For $\mathcal{R}\sim0$ ({\it Path 1}), $\rho_{\alpha_\pm}$ almost overlap each other and ${\rm Var}(X)$ is enough to capture the diffusive nature of the system [see Fig.~\ref{x-segregation}~(c)]. Not only the case of $\mathcal{A}_{\max}$ but also the other case of $\mathcal{R}>0$ ({\it Path 2}) shows that ${\rm Var}(X)\sim t^2$ because the distance between probability density peaks increases almost linearly over time $t$, implying that the ballistic motion of ${\rm Var}(X)$ is more dominated than diffusion [see Fig.~\ref{diffusion-pattern}~(b) and (d)]. To investigate the diffusive nature of the peak, we must consider only the variance of the single peak. Fortunately, we empirically find that the median of $X$ almost separates the peak of the rich side and the others [see Fig.~\ref{x-segregation}~(d)]. Thus, by investigating ${\rm Var}(X_{\rm rich})$ where $X_{\rm rich}=\{X|X>{\rm med}(X)\}$, we can identify the diffusive nature of the peak. We numerically estimate the diffusion exponent $a$ in terms of linear regression for large $t$, and find that for ${\rm Var}(X_{\rm rich})$, $0.5<a<2$ [see Fig.~\ref{diffusion-pattern}~(c) and (d)]. In short, the configurational property $\mathcal{A}$ controls the diffusive nature of the system, and the system lies in the sub-diffusion regime to the super-diffusion regime for large $t$.

\section{HBM model in Watts-Strogatz network}

In this section, we provide detailed numerical simulations for the HBM model in a Watts-Strogatz (WS) network, where
we use the WS network with mean degree $k=4$, not $k=2$. We note that the rewiring procedure with $k=2$ makes several divided components where the income dynamics is not consistent, whereas the case of $k=4$ does.

Since the WS network is not a ring topology, $\mathcal{R}$ is no longer valid. However, $\mathcal{A}$ is still valid for an arbitrary network. The configuration property $\mathcal{A}$ depends on the initial configuration and the rewiring probability $p$ of the WS network. We use two initial configurations: (1) Alternatively allocated $\mathcal{A}_1$ and (2) fully segregated $\mathcal{A}_2$, which correspond to $\mathcal{A}_{\min}$ and $\mathcal{A}_{\max}$ for the 1D ring case, respectively. $\mathcal{A}_1^{(p)}$ ($\mathcal{A}_2^{(p)}$) as an assortativity after the rewiring process with probability $p$ starts from the initial configuration $\mathcal{A}_1$ ($\mathcal{A}_2$).
For the case of $\mathcal{A}_1^{(p)}$, $\mathcal{A}_1^{(0)}=0$ because the number of homogeneous and heterogeneous links is the same. The additional rewiring with $p$ does not change the number of two types of links on average, so that $\mathcal{A}_1^{(p)}\approx0$ [see Fig.~\ref{HBM-WS1}~(a)]. 
For the case of $\mathcal{A}_2^{(p)}$, $\mathcal{A}_{2}^{(0)}=+1-12/(2N)$ because the number of heterogeneous links is $6$ and the total number of links is $2N$. For this case, the additional rewiring with $p$ changes assortativity because the number of homogeneous links is proportional to $(p/2)$, finally, the number of two types of links on average is balanced at $p=1$. Thus, $\mathcal{A}_2^{(p)}\approx1-p$ [see Fig.~\ref{HBM-WS1}~(a)].
\begin{figure}[]
    \includegraphics[width=\textwidth]{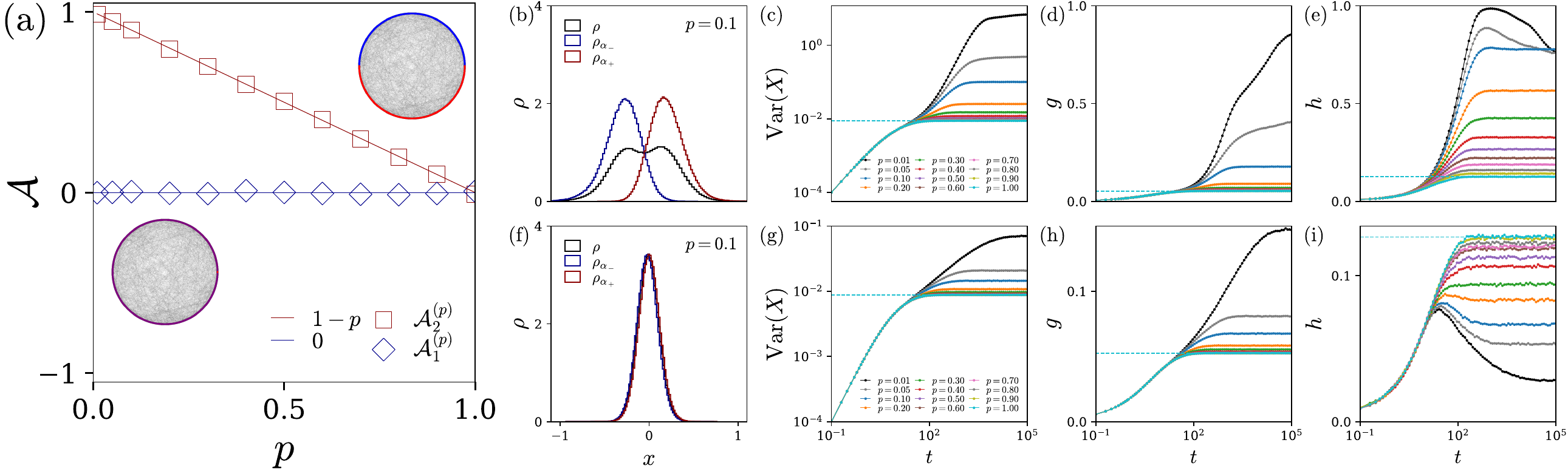}
    \caption{(a) Configuration properties of $\mathcal{A}_{1}^{(p)}$ and $\mathcal{A}_{2}^{(p)}$ with two insets that show two network visualizations for $\mathcal{A}_1^{(p)}$ (upper) and $\mathcal{A}_2^{(p)}$ (lower) with $p=0.1$, respectively. Log-income distribution $\rho(x)$ at $t=10^4$ with $p=0.1$: (b) for $\mathcal{A}_2^{(p)}$ and (c) for $\mathcal{A}_1^{(p)}$. (c)-(e) and (g)-(i) show ${\rm Var}(X)$, $g$, and $h$ for $\mathcal{A}_{1}^{(p)}$ and $\mathcal{A}_2^{(p)}$, respectively. For all cases, $\alpha=10^{-2}, \Delta\alpha=10^{-3}, \beta^2=10^{-3}, J=10^{-1}$ and all data are averaged over 128 ensembles.}
    \label{HBM-WS1}
\end{figure}

In the BM model study by Souma {\it et al}.~\cite{souma2001small}, the small-world (SW) effect changes $\rho(x)$ from log-normal to power-law, and reduces the Gini index $g$. In contrast to the 1D case, in the WS network with sufficiently large $p$, ${\rm Var}(X)$ saturates for large $t$, implying that $\rho(x)$ converges to the stationary distribution [see Figs.~\ref{vcc}~(a), ~\ref{diffusion-pattern}~(a), and ~\ref{HBM-WS2}~(c), (g)]. The HBM model shows a stationary distribution with sufficiently large $p$ and a shift from log-normal to stationary power-law as $p$ increases. However, in the HBM model, both $\mathcal{A}$ and $p$ determine $\rho(x)$, which  is differently dependent on $\mathcal{A}$ even for the same $p$ [see Fig.~\ref{HBM-WS1}~(b) and (f)], and on $p$ even for the same $\mathcal{A}$ [see Fig.~\ref{HBM-WS1}~(g)-(i)]. In particular, $\mathcal{A}_1^{(p)}$ exhibits a stationary unimodal distribution; however, $\mathcal{A}_2^{(p)}$ with appropriate $p$ exhibits a stationary bimodal distribution,  corresponding to the second era of the history of global inequality, reported by Milanovic~\cite{milanovic2024three}. For the \textit{``small-networkness"}, we here consider $p\in[10^{-4},10^{-1}]$, not $p\in[10^{-1},1]$ in the main text.
\begin{figure}[b]
    \includegraphics[width=\textwidth]{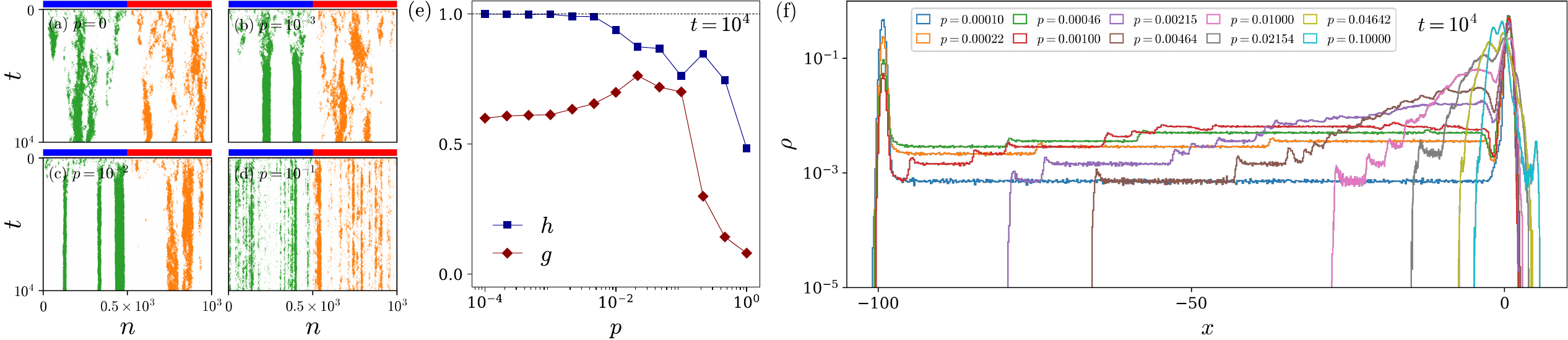}
    \caption{SW effect on HBM model: (a)-(d) Spatiotemporal patterns of top-rich/bottom-poor 10\% (orange/green) class for $\mathcal{A}_2^{(p)}$ for $p=\{0, 10^{-3},10^{-2}, 10^{-1}\}$. (e) Gini index $g$ and Hellinger distance $h$ against $p\in[10^{-4},10^{-1}]$. (f) $\rho(x)$ against log-income $x$ for $p\in[10^{-4},10^{-1}]$. For all cases, $\alpha=10^{-2},\Delta\alpha=5\times10^{-3},\beta^2=10^{-3},J=10^{-1}$, and all data averaged over 128 ensembles  at $t=10^4$.}
    \label{HBM-WS2}
\end{figure}
\begin{figure}[]
    \includegraphics[width=\textwidth]{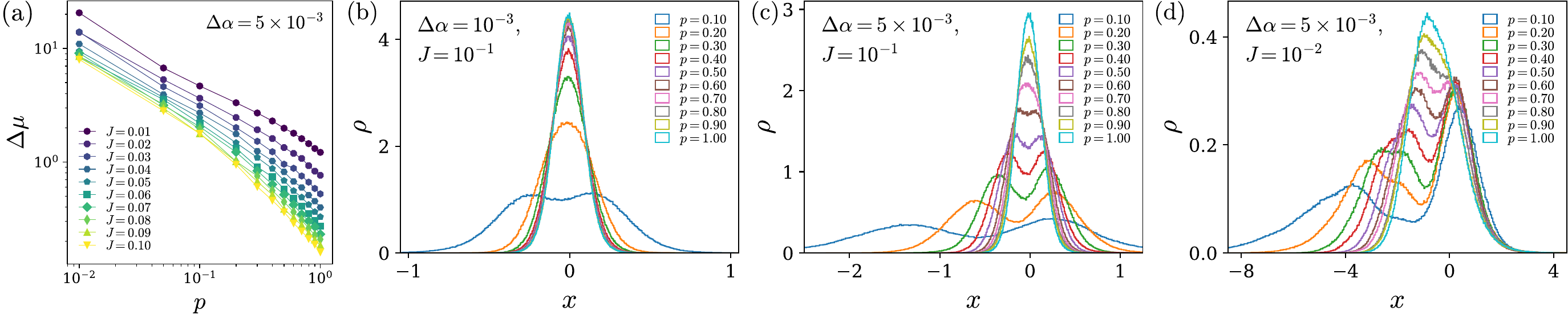}
    \caption{Parameter dependence on HBM model in WS network: (a) $J$ dependence on $\Delta\mu$ against $p$ for $\Delta\alpha=5\times 10^{-3}$ and $J=\{0.01, 0.02,\cdots, 0.10\}$ from top to bottom. (b)-(d) $(\Delta\alpha, J)$ dependence on log-income distribution $\rho(x)$ against $x$: (b) for $(\Delta\alpha,J)=(10^{-3},10^{-1})$, (c) for $(\Delta\alpha,J)=(5\times10^{-3},10^{-1})$, and (d) for $(\Delta\alpha,J)=(5\times10^{-3},10^{-2})$. For all cases, $N=10^4,\alpha=10^{-2},\beta^2=10^{-3}$ and all data are averaged over 128 ensembles at $t=10^4$.}
    \label{HBM-WS3}
\end{figure}

In the limit of $p\to0$, the WS network becomes regular, so that the field exponent $\eta$ is almost the same as that in the 1D BM model [see Fig.~\ref{eta}~(c)]. Thus, the spatiotemporal patterns of income dynamics are also almost the same [see Fig.~\ref{HBM-WS2}~(a) and Fig.~1~(c) in the main text]. For $10^{-4}<p<10^{-2}$, $h\sim1$ represents that $\rho_{\alpha_\pm}$ are almost perfectly decoupled, and $g>1/2$. It is because the between-inequality in $\alpha_{\pm}$ groups guarantees half of the Gini index $g$. In this region, the larger $p$, the larger $g$ because $p$ makes heterogeneous links change $\rho_{\alpha_\pm}$ more diffusive, so that the within-inequality increases. However, for $p>10^{-2}$, both the segregation of $\alpha_\pm$ groups and the variance for each peak decrease, so that both between- and within-inequalities decrease, represented by a rapid decrease in $g$ [see Fig.~\ref{HBM-WS2}~(e)]. For the case of the WS network, in contrast to the the 1D case, $\Delta\mu$ depends not only on $\Delta\alpha$ but also on $J$ [see Eq.~\eqref{dmu} and Fig.~\ref{HBM-WS3}].

As a final remark, we briefly address that the Newman-Watts (NW) model also yields almost similar results as those by the WS network since the stationarity and the shape of distributions mainly depend on the average shortest path length in such a network. Moreover, we confirm that the NW network with the same parameters as before shows a stationary bimodal log-income distribution, where other results are also consistent with the case of the WS network.

\section{Physical interpretations of our study}

In this section, we highlight the conceptual advance in our study related to the universality of local interactions.

For the 1D BM model, the continuum limit of the interaction term in Eq.~(A1)
becomes a spatial diffusion term (or Laplacian). In that case, the equation becomes a 1D stochastic heat equation (SHE) with multiplicative noise. It is known that the Cole-Hopf transformation, which takes the logarithm, gives the 1D Kardar-Parisi-Zhang (KPZ) equation. Thus, our log-income ($x\equiv\ln{c}$) of each region can be considered as a KPZ height field. Also, we have shown that if fluctuation $\beta$ is small, multiplicative noise can be transformed by additive noise, meaning that log-income can be considered as a height field of the Edwards-Wilkinson (EW) equation.

For the HBM model, we introduced heterogeneous growth rates $\alpha_n\in\{\alpha-\Delta\alpha,\alpha+\Delta\alpha\}$ as a quenched disorder, and we still consider space-time noise represented by the Wiener process of each region. Therefore, our model can be mapped into a quenched KPZ equation with space-time noise. 

\begin{figure}[h]
\includegraphics[width=\textwidth]{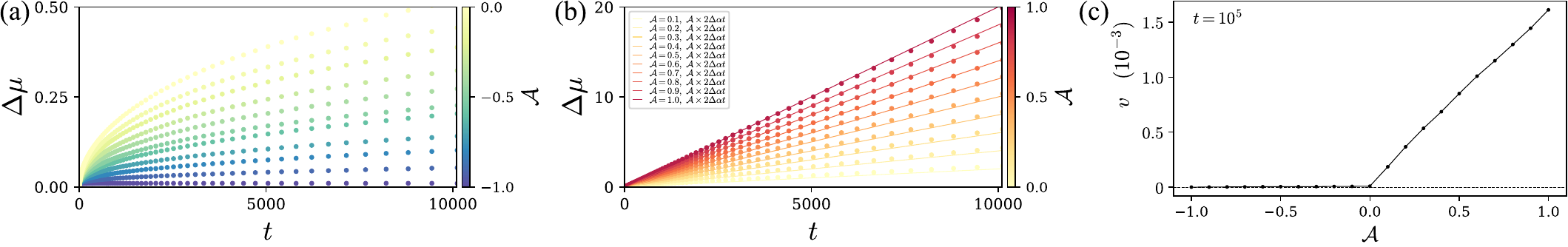}
    \caption{(a), (b) Income level segregation $\Delta\mu$ and (c) average segregation velocity $v$.}
    \label{HBM-rebuttal-1}
\end{figure}

In contrast to quenched 1D-KPZ and 1D-EW equations, the degree of driving force $F$ is not an control parameter since growth rates $\alpha_n$ are given and fixed. The important thing is that in our case, configurational properties play as control parameters for a Pinning-Depinning (PD)-like transition. As we have shown in Fig.~\ref{x-segregation}, income level segregation $\Delta\mu$ sub-linearly increases for \textit{Path 1} configurations but linearly increases for \textit{Path 2} configurations. If we define average segregation velocity $v\equiv\Delta\mu/t$, $v$ shows continuous phase transition over $\mathcal{A}$ [see Fig.~\ref{HBM-rebuttal-1}-(c)].
\begin{align}
    v\equiv\Delta\mu/t\sim
    \begin{cases}
        0 & \text{for {\it Path 1}: $(\mathcal{A}\leq0,\mathcal{R}\sim0)$}, \\
        \mathcal{A}\times 2\Delta\alpha & \text{for {\it Path 2}: $(\mathcal{A}>0,\mathcal{R}>0)$}.
    \end{cases}
    \label{v}
\end{align}
It shows that height segregation of the surface emerges from the configurational properties of growth rates.

Since we simultaneously consider quenched disorder and space-time noise, there are different growth patterns from the 1D-KPZ or 1D-EW equation. As we have shown in Fig.~\ref{diffusion-pattern}, various configurational properties give different diffusion exponents (twice the growth exponent in the surface growth model). It means that the coexisting interplay of quenched disorder and space-time noise makes an unseen growth exponent from the quenched KPZ ($1/3$) and EW ($1/4$) equations, and that is the meaning of our terminology ``anomaly''.

\section{Spectrum of growth rates}

Finally, in this section, we discuss the  effect of the growth-rate spectrum on our model study.  

The configurational property $\mathcal{A}$ is well defined even if we consider a continuous spectrum of growth rates $\alpha$. Also, if we take the weighted Kuramoto oscillator's order parameters, $\mathcal{R}$, is still defined in the 1D case.
The presence or absence of bimodality depends heavily on the proportion of the population corresponding to the growth rates (because even if the income levels of the distributions are significantly different, the height difference may be so large that the smaller distribution appears to be absorbed by the larger distribution). Hence, the second era of bimodality in global income inequality can be seen as a result of an appropriate distribution of income levels and population shares between developed and developing countries. However, the important thing is the regional segregation of income level is consistent with our result even if we take continuous spectrum of growth rates such as uniform distribution $\alpha_n\in[\alpha-\Delta\alpha,\alpha+\Delta\alpha]$ or normal distribution $\alpha_n\sim\mathcal{N}(\mu,\sigma^2)$ and it matches the concept of \textit{location-based inequality}~\cite{milanovic2015global}.

Since normally distributed growth rates in the 1D ring case has been tested in our previous study~\cite{hur2024interplay}, it was intentionally omitted. Instead, we focus on clarifying the analytics of the binary case in detail. The analysis of the aggregate growth rate, Gini index, and correlations between growth rate and log-income through the configurational property $\mathcal{A}$ under a normally distributed growth rate, see the Supplemental Material (SM) of our previous paper~\cite{hur2024interplay}.
\end{widetext}

\bibliographystyle{apsrev4-2}
\bibliography{ref-HBM_updated}

\begin{thebibliography}{59}
\bibitem [{Pea()}]{Pearson}%
  \BibitemOpen
  \href@noop {} {}\bibinfo {note} {The value of the Pearson correlation coefficient is bounded between -1 and 1.}\BibitemShut {Stop}%
\bibitem [{log()}]{log-income}%
  \BibitemOpen
  \href@noop {} {}\bibinfo {note} {For the BM model, $\rho(C)$ does not converge over time. However, for normalized income $c\equiv C/\bar{C}$, $\rho(c)$ converges over time such conditions that mean-field case and dense networks, and gives more mathematical conveniences. The log-normalized income is defined as $x\equiv\ln{c}$.}\BibitemShut {Stop}%
\bibitem [{SDE()}]{SDE-normalizedwealth}%
  \BibitemOpen
  \href@noop {} {}\bibinfo {note} {By using $\bar{C}=C(0)e^{\alpha t}$, we get $dc_n=d(C_n/\bar{C})=dC_n/\bar{C}-(C/\bar{C}^2)d\bar{C}=dC_n/\bar{C}-\alpha c_ndt$ in Eq. (2).}\BibitemShut {Stop}%
\bibitem [{OU()}]{OU}%
  \BibitemOpen
  \href@noop {} {}\bibinfo {note} {In the ordinary OU process, $\rho^{(\rm OU)}_{\rm eq}(x)$ is a Gaussian with variance $\sigma^2$ that converges to the ratio of the diffusion coefficient to the dissipation factor, $\sigma^2\to\beta^2/[2J\eta\theta(\eta)]$.}\BibitemShut {Stop}%
\end{thebibliography}
\end{document}